\documentclass{aa}  
\usepackage{comment}
\usepackage{threeparttable}
\usepackage{caption}
\usepackage{booktabs}
\usepackage{subcaption}
\usepackage{tikz}
\usepackage{natbib}
\usepackage{amssymb}
\usepackage{placeins}
\usepackage{stfloats}
\usepackage{graphicx}
\usepackage{hyperref}
\usepackage{float}
\usepackage{tabularx}
\usepackage{array}
\usepackage{siunitx}
\usepackage{textcomp}

\usepackage{txfonts}

\begin{document}

   \title{The RoPES project with HARPS and HARPS-N \thanks{\email{nicola.nari@lightbridges.es}}}
   \subtitle{III. Two candidate planets orbiting the G-type star HD~161098}
   
   \author{N.~Nari \inst{\ref{lbridges},\ref{iac},\ref{ull}}
\and
A.~Su\'arez Mascare\~no \inst{\ref{iac},\ref{ull}}
\and
J.\,I.~Gonz\'alez Hern\'andez \inst{\ref{iac},\ref{ull}}
\and
A.\,K.~Stefanov \inst{\ref{iac},\ref{ull}}
\and
R.~Rebolo \inst{\ref{iac},\ref{ull},\ref{csic}}
\and 
J.\,M.~Mestre \inst{\ref{unipd}}
\and
X.~Dumusque \inst{\ref{unige}}
\and
M.~Cretignier \inst{\ref{unioxf}}
\and 
V.\,M.~Passegger \inst{\ref{subaru},\ref{iac},\ref{ull},\ref{sw}}
\and
L.~Mignon \inst{\ref{unigre},\ref{unige}}
\and 
F.~Manni \inst{\ref{torver},\ref{oato}}
\and 
R.\,G.\,S.\,B.~De~Amorim \inst{\ref{rgdn_natal}}
}
   \institute{
Light Bridges S.L., Observatorio del Teide, Carretera del Observatorio s/n, Güimar, 38500 Tenerife, Spain \label{lbridges}
\and
Instituto de Astrof{\'i}sica de Canarias, 38205 La Laguna, Tenerife, Spain \label{iac}
\and
Departamento de Astrof{\'i}sica, Universidad de La Laguna, 38206 La Laguna, Tenerife, Spain \label{ull}
\and
Consejo Superior de Investigaciones Cient\'{i}ficas, Spain \label{csic}
\and
Dipartimento di Fisica e Astronomia ``Galileo Galilei''
, Università di Padova, Vicolo dell’Osservatorio 3, 35122 Padova, Italy \label{unipd}
\and
Observatoire de Gen\`eve, D\'epartement d'Astronomie, Universit\'e de Genève, Chemin Pegasi 51b, 1290 Versoix, Switzerland \label{unige}
\and 
Department of Physics, University of Oxford, Oxford OX13RH, United Kingdom \label{unioxf}
\and
Subaru Telescope, National Astronomical Observatory of Japan, 650 N Aohoku Place, Hilo, HI 96720, USA \label{subaru}
\and
Hamburger Sternwarte, Gojenbergsweg 112, 21029 Hamburg, Germany \label{sw}
\and
Universit\'e Grenoble Alpes, CNRS, IPAG, 38000 Grenoble, France \label{unigre}
\and 
Dipartimento di Fisica, Università di Roma “Tor Vergata”, Via della Ricerca Scientifica 1, 00133 Rome, Italy \label{torver}
\and
INAF – Osservatorio Astrofisico di Torino, via Osservatorio 20, 10025 Pino Torinese, Italy
\label{oato}
\and
Departamento de Física Teórica e Experimental,  
Universidade Federal do Rio Grande do Norte,  
Campus Universitário, Natal, Rio Grande do Norte, 59072-970, Brazil \label{rgdn_natal}
}
\date{Received 23 January 2026 / Accepted 09 March 2026}

  \abstract 
  {The development of refined instruments and techniques for a detailed analysis of the radial velocities (RVs) of stars other than the Sun allows the presence of planetary signals of amplitude below 1\,m\,s$^{-1}$ to be investigated. Long-term RV surveys allow the detection of Earth-like and super-Earth-like planets in the habitable zones of Sun-like stars, prime targets for future missions for the atmospheric characterization of exoplanets.}
   {We present the analysis of the nearby G8\,V-type star HD 161098 ($V$ = 7.68\,mag, $d$ =\,29.75 pc). We searched for terrestrial planets in the habitable zone.}
   {We combined historical datasets with new data collected in an ongoing blind search program with HARPS, HARPS-N, and ESPRESSO. 
   We utilized recently developed tools to extract RVs and to deal with the analysis of stellar activity. We performed a joint analysis of RVs and activity indicators to separate the planetary signals from those related to activity.}
  {We detected two sub-m\,s$^{-1}$ signals that we claim as candidate planets. We are not able to confirm their nature with certainty. Candidate HD 161098 b has an orbital period of 72.578$_{-0.060}^{+0.059}$\,\si{\day} and a minimum mass of 3.63 $\pm$ 0.59\,M$_\oplus$. HD 161098 c has an orbital period of 682.5$_{-9.9}^{+9.5}$\,\si{\day} and a minimum mass of 7.8$_{-1.4}^{+1.5}$\,M$_\oplus$.If confirmed, candidate HD 161098 c would reside in the optimistic habitable zone of the star.
  We find a magnetic cycle of 4090$_{-130}^{+140}$\,\si{\day} period and a rotation period of 28.22$_{-0.35}^{+0.30}$\,\si{\day}.
  Our analysis sets the stage for future observing campaigns of the star, finalized for the confirmation of our results.
  }
  {We are entering the sub-m\,s$^{-1}$ era at long orbital periods with a combination of stellar activity treatment and long-term campaigns.}


   \maketitle
   
\section{Introduction}
The first discovery of an exoplanet orbiting a main-sequence star \citep{1995_mayor_first_planet} was made with the radial velocity (RV) method anticipated by \citet{struve_1952} (a review is available by \citealp{2025_rv_method}).
Since then, more than 6,000 planets have been discovered, as reported in the  NASA Exoplanet Archive\footnote{\url{https://exoplanetarchive.ipac.caltech.edu/}} \citep{2025_christiansen_archive}. 
Even though detections via the transit method \citep{2000_transit_henry,2000_transit_charbonneau} outnumbered those found via the RV method, the RV method remains the most fruitful for the  mass determination of exoplanets. 

A high-resolution spectrograph is necessary to achieve the required RV precision for detecting exoplanets.
The High Accuracy Radial velocity Planet Searcher (HARPS; \citealp{2003_mayor_harps}), installed in 2003 at 3.6\,m telescope at the Observatory of La Silla, Chile, was the first instrument able to reach an RV precision below 1\,m\,s$^{-1}$. 
HARPS-N \citep{2012_cosentino_harpsn,2014_harpsn_cosentino}, an improved version of HARPS, was installed at the Telescopio Nazionale Galileo at the Observatorio del Roque de Los Muchachos, Spain.  

Long-term RV surveys enable the detection of planets on wide orbits. 
The ROcky Planets in Equatorial Stars (RoPES) program \citep{2018_ropes_alejandro} is a long-term survey to investigate a sample of 17 G and K dwarf-type stars with HARPS and HARPS-N. 
The main aim of the project is to detect Earth-like and super-Earth-like planets in the habitable zone (HZ) of stars similar to the Sun, and to provide targets for future atmospheric characterization with facilities such as ArmazoNes high Dispersion Echelle Spectrograph \citep{2022_marconi_andes,2023_palle_andes}, Habitable World Observatory \citep{2026_hwo_feinberg}, and LIFE \citep{2022_life_quanz}. To achieve this goal, RoPES combines historical data and recent observations from the program itself.

For HD 161098, one of the stars in the program, we merged the HARPS and HARPS-N datasets with ESPRESSO observations from a recent observing campaign. ESPRESSO is the most precise spectrograph in the world, with a precision on RVs up to 10\,cm\,s$^{-1}$ \citep{2021_espresso_pepe}. It is installed at the Very Large Telescope at the Paranal Observatory, Chile.

The paper is structured as follows. In Sect. \ref{sec_obs}, we discuss the observations we have used in our analysis. In Sect. \ref{sec_stellar_parameters}, we describe the star HD 161098. In Sect. \ref{sec_analysis}, we discuss the analysis on the system. In Sect. \ref{sec_dis}, we discuss the results of our analysis. In Sect. \ref{sec_con}, we summarize our work.
\section{Observations}
\label{sec_obs}
\subsection{HARPS and HARPS-N}
HARPS and HARPS-N are fiber-fed, high-resolution spectrographs. They span wavelengths between 380\,\si{\nano\meter} and 690\,\si{\nano\meter}. The instruments have a resolving power of $\sim$ 115\,000. They are both pressure- and temperature-stabilized to minimize the drift in RV.
Wavelength-calibrated, high-resolution spectra are provided by the data reduction software ({\tt DRS}). For HARPS, we used version 3.3.6 of the {\tt DRS} adapted from the \mbox{ESPRESSO {\tt DRS} \footnote{\url{https:/www.eso.org/sci/software/pipe_aem_main.html}}}, which corrects for various systematics, such as lamp aging and lamp changes, thus increasing the consistency of long-term surveys. We used version 3.0.1 of the {\tt DRS} for HARPS-N.
Both {\tt DRS}s provide RVs calculated with the cross-correlation function (CCF) technique \citep{1953_fellgett_ccf}. Other by-products of the CCF are the full width at half maximum (FWHM) and the bisector time span (BIS), among others. 

HARPS underwent a fiber link update in June 2015 \citep{2015_lo_curto_fiber}, which introduced an RV offset. We considered observations collected before that date to be the H03 dataset, and observations collected after that date the H15.
For each of these datasets, we fitted a zero-point and a jitter term. More details on the methods adopted in our analysis are reported in Appendix \ref{methods_sec}. We refer to the HARPS-N dataset as HN.

We extracted HARPS and HARPS-N CCF RVs with {\tt YARARA} \citep{2021_yarara_cretignier,2023_yarara_cretignier}. {\tt YARARA} corrects for different systematics at the spectral level, such as cosmic rays, interference patterns, telluric lines, point spread function variability, ghosts, stitching of the detector, and thorium-argon lamp contamination. 
{\tt YARARA} checks for anomalous CCF, anomalous S/N, or anomalous residuals in the spectra. Observations that do not surpass a certain threshold are not taken into account for weighting the recipes of the pipeline, and we discarded them from the analysis for this reason. We had a total of 339 HARPS and HARPS-N epochs, split between 128 H03 epochs, 101 H15 epochs, and 110 HN epochs. After the quality control of {\tt YARARA}, we remained with 119 H03 epochs, 80 H15 epochs, and 102 HN epochs. 
We have a standard deviations of the different datasets of 2.17\,m\,s$^{-1}$ for H03, 1.41\,m\,s$^{-1}$ for H15, and 1.51\,m\,s$^{-1}$ for HN. 
The mean error per dataset is 0.44\,m\,s$^{-1}$ for H03, 0.51\,m\,s$^{-1}$ for H15, and 0.33\,m\,s$^{-1}$ for HN on nightly binned observations. For H03, we collected single exposures of 900\,\si{\second}. We have a mean S/N on the single exposure of 176 for H03 at order 55.
For H15, we collected three consecutive exposures of 300\,\si{\second} each, and we have a mean S/N of 93 at order 55 for a single exposure. For HN we collected three consecutive exposures of 300\,\si{\second} each with a mean S/N of 113 at order 55.

\subsection{ESPRESSO}
ESPRESSO \citep{2021_espresso_pepe} is a high-resolution spectrograph. The instrument covers a wavelength range comprised between 378.2\,\si{\nano\meter} and 788.7\,\si{\nano\meter}. For our observations, we used the high-resolution 1-UT mode, which has a resolving power of $\sim$ 140\,000. ESPRESSO is contained in a vacuum vessel to prevent temperature and pressure shifts.
We used {\tt DRS} version 3.3.10 to extract CCF RVs and activity indicators. Hereafter, we refer to ESPRESSO observations as E19. We have 51 epochs of observations. 
The standard deviation of the E19 dataset is 1.28\,m\,s$^{-1}$ for nightly binned observations. The mean error is 0.16\,m\,s$^{-1}$ on nightly binned observations. 
We collected three consecutive exposures per night, with an exposure time of 300\,\si{\second} per exposure, with a mean S/N of 203 at order 145. 
\section{HD 161098: Stellar parameters}
\label{sec_stellar_parameters}
HD 161098 is a bright ($V$ = 7.67 $\pm$ 0.01 mag; \citealp{2000_tycho_catalogue}) G8\,V star, close to the Solar System ($d$ = 29.75 $\pm$ 0.02\,pc; \citealp{2020_gaia_edr3}). 
HD 161098 has an effective temperature of 5610 $\pm$ 50\,K, with a mass of 0.837 $\pm$ 0.029\,M$_\odot$, a radius of 0.866$_{-0.020}^{+0.020}$\,R$_\odot$, and a luminosity of 0.6769 $\pm$ 0.0015\,L$_\odot$.
In Table \ref{tab_stel_par} we present a summary of the main characteristics of the star.

We used the method described by \citet{kopparapu_habitable_zone} to calculate the boundaries of the HZ in the case of a 1\,M$_\oplus$ planet.
We considered their recent Venus and early Mars regimes respectively as the inner and outer edges for the optimistic HZì. We considered the runaway and maximum greenhouse regimes as the conservative inner and outer edges of the HZ. We found the conservative HZ between 0.8155 $\pm$ 0.0025\,au and 1.4097$_{-0.0062}^{+0.0063}$\,au, which corresponds to an orbital period between 294.0$_{-5.1}^{+5.4}$\,\si{\day} and 668$_{-12}^{+13}$\,\si{\day} for a circular orbit. We found the optimistic HZ comprised between 0.6216 $\pm$ 0.0014\,au and 1.4808 $\pm$ 0.0066\,au, which corresponds to an orbital period between 195.6$_{-3.4}^{+3.5}$\,\si{\day} and 719$_{-13}^{+14}$\,\si{\day} for a circular orbit.

\section{Analysis}
\label{sec_analysis}
\begin{figure*}[!htbp]
    \centering
    \includegraphics[width=\textwidth]{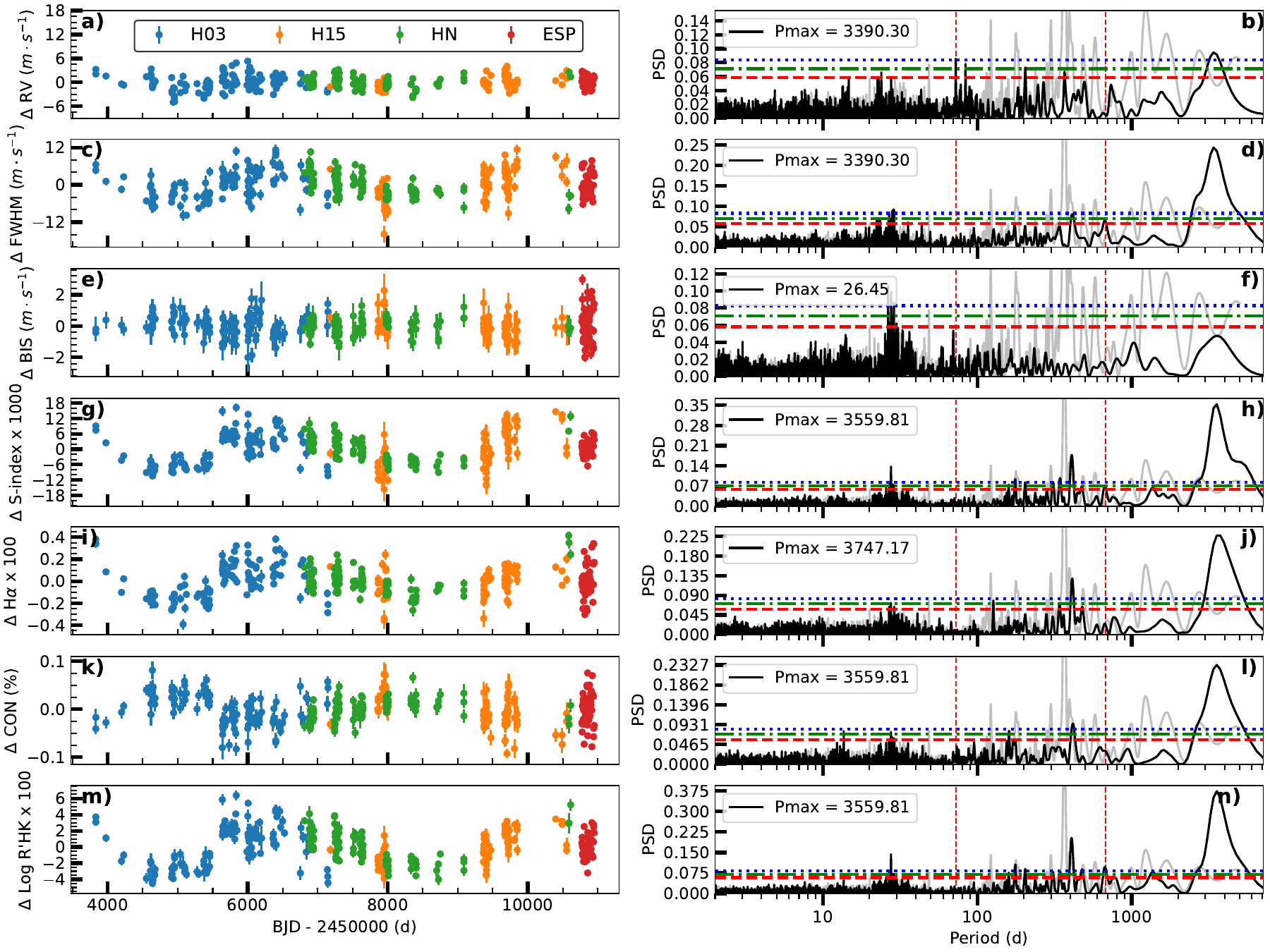}
    \caption{Time series of RV and activity indicators (left) and their respective GLS periodograms (right). The dashed red line is the 10 \% false alarm probability (FAP), the dashed-dotted green line is the 1 \% FAP, and the dotted blue line is the 0.1 \% FAP. We plot dashed lines at the periods of the candidate planets. Panel (a): RV time series. Panel (b): RV GLS periodogram. Panel (c): FWHM time series. Panel (d): FWHM GLS periodogram. Panel (e): BIS time series. Panel (f): BIS GLS periodogram. Panel (g): $S$ index time series. Panel (h): $S$ index GLS periodogram. Panel (i): H$\alpha$ time series. Panel (j): H$\alpha$ GLS periodogram. Panel (k): CON time series. Panel (l): CON GLS periodogram. Panel (m): $\log{R'_{\rm HK}}$ time series. Panel (n): $\log{R'_{\rm HK}}$ GLS periodogram. The window function of the observations is shown in light gray.
    The vertical red lines in the periodograms indicate the period of the candidate planets we found in the analysis. The candidate at 72.5\,\si{\day} only appears in RV.}
    \label{hd161098_activity_indicators}
\end{figure*}

An analysis of stellar activity is presented in Sect. \ref{stellar_activity_subsec}. We applied a multidimensional Gaussian process (GP) (Sect.~\ref{stellar_activity_appendix}) analysis to simultaneously fit for activity indicators and RVs \citep{2015_rajpaul_multigp,2023_barragan_multigp}. Multidimensional GPs are less likely to overfit the time series, and especially preserve the long-period signals when compared to 1D GP \citep{2026_nari_hd176986}.
In Fig. \ref{hd161098_activity_indicators} we show the time series used in this analysis and the corresponding GLS periodograms \citep{2009_gls_zeichmeister}.

\subsection{Stellar activity}
\label{stellar_activity_subsec}
We used different activity proxies: the full width at half maximum of the CCF, directly provided by the {\tt DRS}; the
$S$ index, related to the intensity of the chromospheric emission in the \ion{Ca}{ii} H \& K lines (we used the method of \citet{2011_lovis_cycle} to calculate the $S$ index); the H$\alpha$, related to the strength of the H$\alpha$ emission line (we calculated the H$\alpha$ index with the method described in\citet{2011_gomes_da_silva_halpha}); the bisector time span, which measures the asymmetry in the profile of a CCF \citep{2001_queloz_bisector}; and the contrast (CON) of the CCF given by the relative ratio between the center of the CCF and the wings.
We see in Fig. \ref{hd161098_activity_indicators} the GLS periodogram of the different activity indicators points toward a magnetic cycle of period $P$ $\sim$ 3500\,\si{\day}.
We modeled the magnetic cycle of the star with two sinusoids, at the period of the cycle and at its first harmonic. We tried different numbers of harmonics, and this solution was the favored one in terms of the Bayesian evidence (Sect. \ref{methods_sec}). We used a uniform prior on the period of the cycle $\mathcal{U}$(2000\,d,6000\,d). We found a significant amplitude for the magnetic cycle in FWHM, $S$ index, H$\alpha$, and CON. We determined slightly different magnetic cycle periods depending on the indicator: $P_{\rm S\,index}$ = 3845$_{-78}^{+80}$\,\si{\day}, $P_{\rm FWHM}$ = 3590 $\pm$ 130\,\si{\day}, $P_{\log{R'_{\rm HK}}}$ = 3900$_{-97}^{+95}$\,\si{\day}, $P_{\rm H-\alpha}$ = 3810$_{-260}^{+280}$\,\si{\day}, and $P_{\rm CON}$ = 3740 $\pm$ 210\,\si{\day}. They are all compatible with each other within 1$\sigma$ except for $P_{\rm FWHM}$ and $P_{\rm S\,index}$, and $P_{\rm FWHM}$ and $P_{\log{R'_{\rm HK}}}$, which are compatible with each other at 1.66$\sigma$ and 1.89$\sigma$, respectively.
We found an amplitude of the main component of the magnetic cycle defined at 14.6$\sigma$ for $S$ index, at 9.9$\sigma$ for FWHM, at 11.3$\sigma$ for $\log{R'_{\rm HK}}$, at 5.4$\sigma$ for H$\alpha$, and 8.6$\sigma$ for CON.

We applied a 1D GP to model the rotation-induced effect of the star. We found $P_{\rm rot}$ = 28.22$_{-0.34}^{+0.28}$\,\si{\day} in $S$ index. This value is compatible with the values found in all the other activity indicators.
We analyzed the activity indicators from the HARPS RVBank \citep{2020_harps_rv_bank_trifonov}. 
Due to the limited number of nights available, it did not add relevant information to the analysis. We do not consider it for the analysis that follows.
\subsection{Planetary signal search}
\label{fip_sect}
\begin{figure}[]
    \begin{minipage}{0.45\textwidth}
        \includegraphics[width=\linewidth]{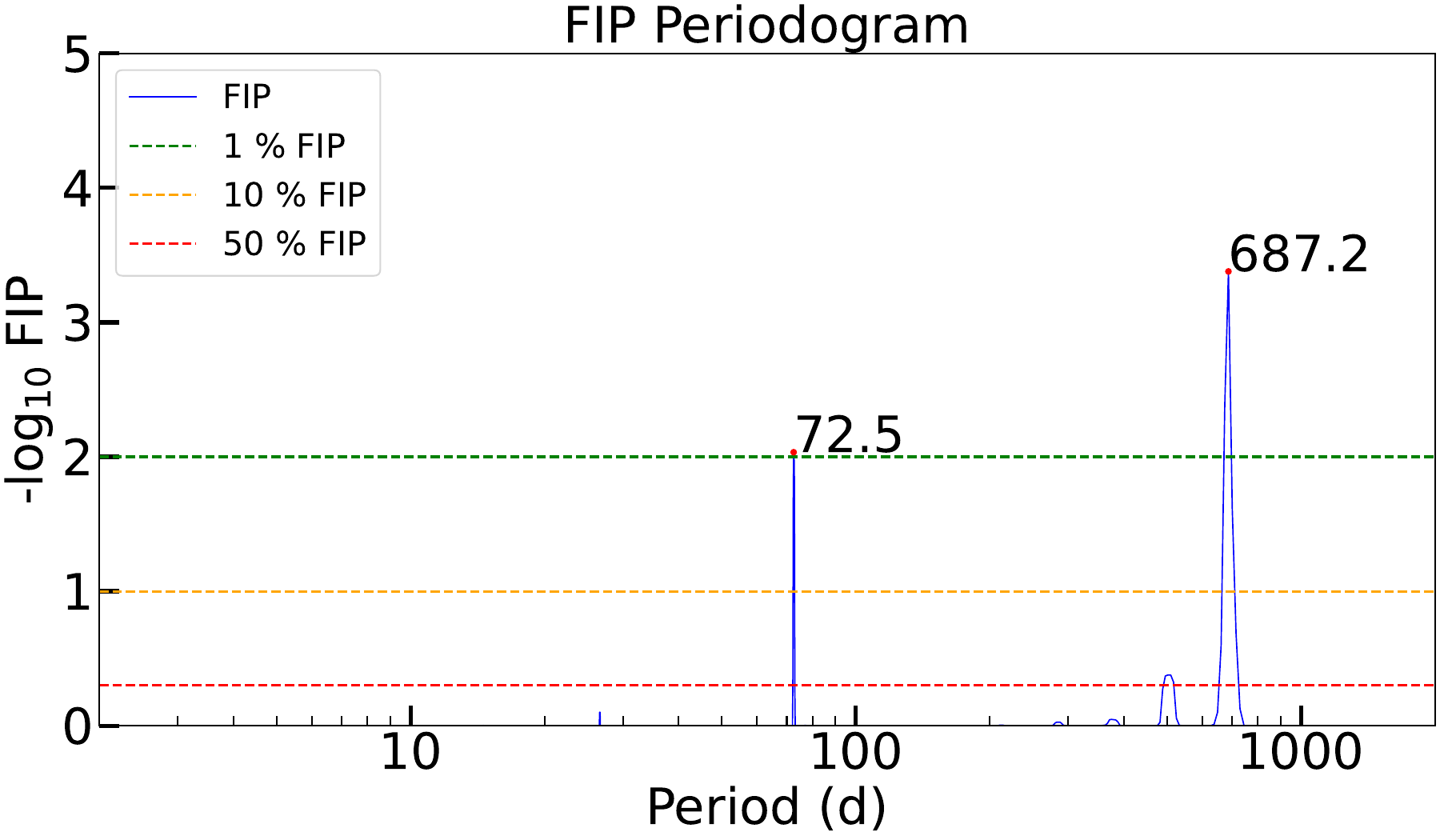}
    \end{minipage}
    \caption{FIP periodogram for HD 161098. The horizontal lines represent the different FIP level of 50\%, 10\%, 1\%. 
    }
    \label{hd161098_fip_periodogram}
\end{figure}

We searched for candidate planetary signals in our RV dataset by measuring a false inclusion probability (FIP) periodogram \citep{2022_fip_hara}. We fitted three sinusoids simultaneously, which shared a log-uniform prior between 2\,\si{\day} and 1000\,\si{\day}. We chose these priors to minimize the risk of fitting for harmonics of the magnetic cycle, which has a period of more than 3500\,\si{\day}.
 We modeled the stellar activity in a multidimensional GP framework (for details on the nested sampling setup, see Sect. \ref{methods_sec}). We used FWHM and $S$ index as ancillary activity indicators because they have the most significant determination of the cycle among independent activity indicators. We considered a magnetic cycle for each of the datasets. We shared the period and phase of the components of the cycle among the RVs and activity indicators. We show in Fig. \ref{hd161098_fip_periodogram} the results of the FIP analysis. We found two periods with FIP below 1 \%, one at 72.5\,\si{\day} and the other at 687.2\,\si{\day}. This is the threshold indicated in \citet{2022_fip_hara} as a detection criterion. We did not find other signals with FIP below 10 \%. The presence of only two signals of interest justifies the usage of a three-sinusoid model.

We also tried an iterative blind addition of planets to the model. 
First, we considered a one-sinusoidal model, along with the multidimensional GP framework with FWHM and $S$ index. We used 
a log uniform prior on the period $\mathcal{U}$(0.6,6.9) corresponding to potential periods between $\sim$ 2\,\si{\day} and $\sim$ 1000\,\si{\day}. We found a sinusoidal signal of amplitude $K$ = 0.68 $\pm$ 0.12\,m\,s$^{-1}$ at a period of 678$_{-13}^{+11}$\,\si{\day}. The one-sinusoidal model is favored in terms of $\Delta$ ln Z by + 7.0 over the flat model ( Sect. \ref{methods_sec}). 

We added a second sinusoid using the same prior on the period of the one-sinusoidal model, $\mathcal{U}$(0.6,6.9), for both sinusoids. We found a $\Delta$ lnZ of +7.0 compared to the one-sinusoidal model. We found a sinusoidal with amplitude $K$ = 0.63 $^{+0.11}_{-0.12}$\,m\,s$^{-1}$ and period $P$ = 676 $^{+11}_{-10}$\,\si{\day} and a sinusoidal with amplitude $K$ = 0.63 $\pm$ 0.10\,m\,s$^{-1}$ and period $P$ = 72.580 $^{+0.058}_{-0.057}$\,\si{\day}.

We performed a three-sinusoidal model to search for additional significant signals. The three-sinusoidal model is disfavored, compared to the two-sinusoidal model, by $\Delta$ lnZ = $-0$.2. We found the two-sinusoidal model to be the best model to explain the data. 
We will call the signal at $\sim$ 72.5\,\si{\day} candidate HD 161098 b, and the signal at $\sim$ 680\,\si{\day} candidate HD 161098 c. 
We performed different tests to verify the planetary nature of the candidates.

\subsection{Stability of signals}

We made a test to check the stability of the signals through the observing campaigns. We took inspiration from the apodized test \citep{2022_hara_apodized}.
We multiplied the amplitude of the sinusoid by a Gaussian term $G$:
\begin{equation}
    \Delta RV = -K \cdot \sin\left(2\pi \cdot \frac{t - t_0}{P_{\text{pl}}}\right) \cdot G(\mu, \sigma)
    \label{eq_gaussian}
\end{equation}
\noindent
Here $\mu$ is the center of the Gaussian, while $\sigma$ is the width of the Gaussian. The planetary signals are stable on a timescale larger than the time span of observations; it is not be possible to define the center of the Gaussian, and the width of the Gaussian should be longer than the baseline of the observations.
We used normal priors, $\mathcal{N}$(72.5d,0.3d) for candidate b and $\mathcal{N}$(680d,10d) for candidate c.

\begin{figure}[]
    \centering
    {\includegraphics[width=\linewidth]{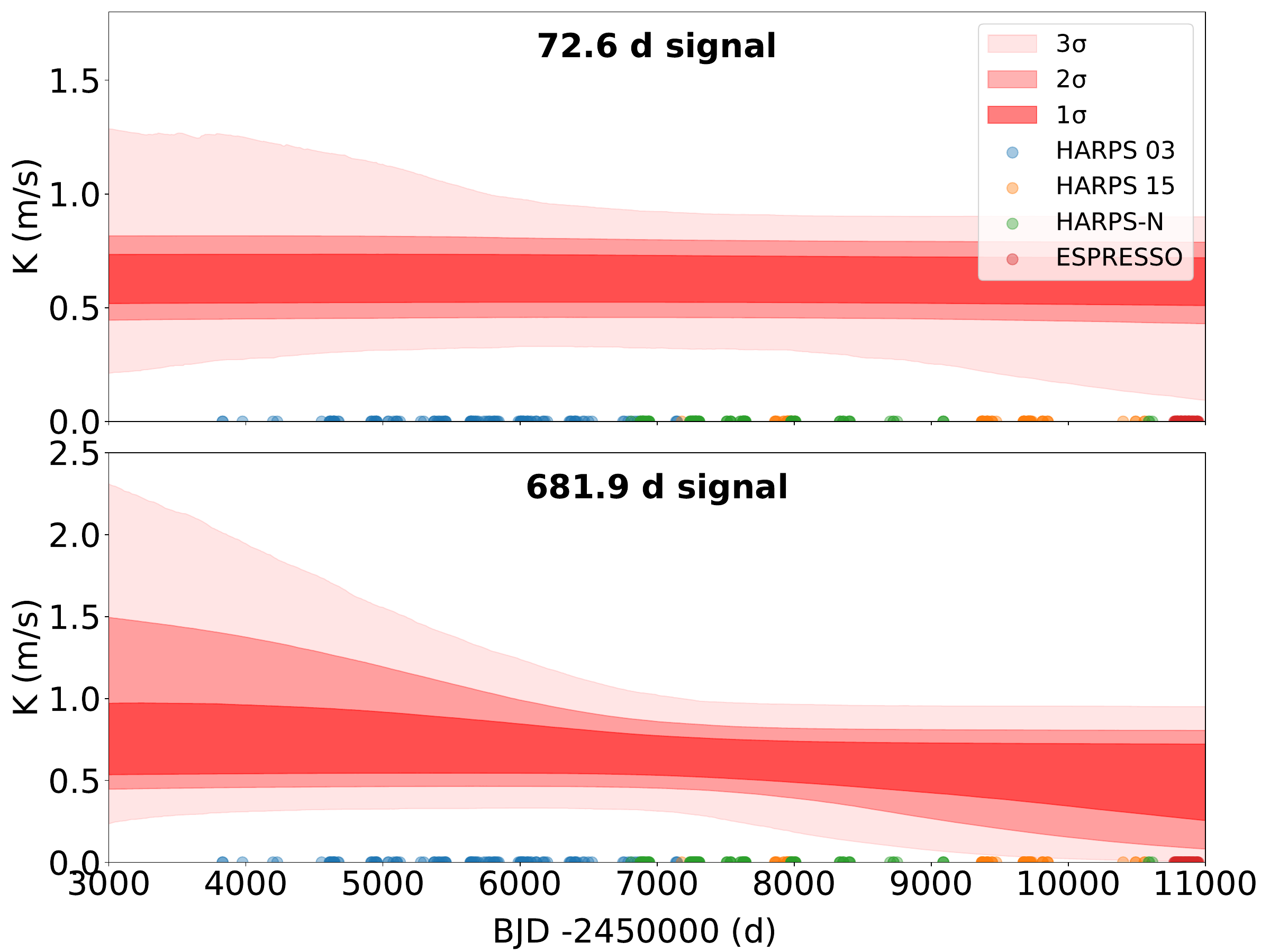}};
    \hfill
    \caption{Apodized test for candidate planets of HD161098.}
    \label{fig_apodized}
\end{figure}

We found for signal b $\mu_b$ = 2455900 $\pm$ 4100 BJD and ln$\sigma_b$ = 14.4$_{-4.0}^{+3.8}$.
We found for signal c $\mu_c$ = 2454500$_{-3100}^{+4400}$ BJD and ln$\sigma_c$ = 12.3$_{-3.8}^{+5.2}$.
The center of the Gaussian is not well defined for both signals. The width of the Gaussian is above or comparable with the baseline of the observations within the error bars. The evidence of the model is worsened by $\Delta$lnZ = $-$1.6 
compared to a model with two sinusoids alone with the same priors on the periods.
We show in Fig. \ref{fig_apodized} the product of the posterior distribution of the amplitudes of the signals multiplied by the Gaussian filters. 

Candidate b is stable through the time series. We found an asymmetric shape of the signal for candidate c, with a preference for the epochs of the H03 dataset. 
We tested an informed model without apodization on the H15, HN, and E19 datasets without H03. We used as priors $\mathcal{N}$(72.5d,5d) for candidate b and $\mathcal{N}$(680d,40d) for candidate c, to also test the period of the candidates.
We found $K$$_b$ = 0.69 $\pm$ 0.11\,m\,s$^{-1}$ and $P_b$ = 72.81 $\pm$ 0.086\,\si{\day}. We found $K$$_c$ = 0.49 $\pm$ 0.13\,m\,s$^{-1}$ and $P_c$ = 692 $\pm$ 19\,\si{\day}. The amplitude of the two signals is within 1$\sigma$ compatible with the amplitudes found in the full analysis. The period of candidate b is different at 3.1$\sigma$, while the period of candidate c is within 1$\sigma$, compatible with the period found in the full dataset. The time of inferior conjunction is BJD 2460256$_{-85}^{+75}$ for candidate b and BJD 2460880.3$\pm$ 2.6 for candidate c. Both are compatible within 1$\sigma$ with the values found in the full analysis. The difference in the period of candidate b could be related to the different sampling of the signal, once we exclude the H03 dataset. 
We repeated the same test with a uniform prior on the period, $\mathcal{U}$(50d,100d) for candidate b and $\mathcal{U}$(500d,1000d) for candidate c. 
We found $K$$_b$ = 0.67 $\pm$ 0.11\,m\,s$^{-1}$ and $P_b$ = 72.81 $_{-0.09}^{+0.013}$\,\si{\day}. We found $K$$_c$ = 0.53 $_{-0.14}^{+0.16}$\,m\,s$^{-1}$ and $P_c$ = 800 $_{-115}^{+32}$\,\si{\day}. The absence of a large portion of the dataset and a baseline that $\sim$ 3000\,\si{\day} shorter can explain the difficulties in detecting the long-period signal, but it could also be related to an artifact introduced by the H03 dataset. Only additional observations can solve the riddle. 
\subsection{Signals in activity indicators}

We searched for the sinusoidal signals we found in RVs in the activity indicators. For simplicity, we used a multidimensional GP model with RVs and two activity indicators. Even so, we fitted the signal only in RV and one activity indicator at a time. We shared the period and phase of the sinusoidal signals. The priors on the periods were $\mathcal{U}$(50d,100d) for candidate HD 161098 b and $\mathcal{U}$(500d,1000d) for candidate HD 161098 c. We summarize the values obtained in Table \ref{table_signal_activity_indicators_rv}.
All the models with the candidate planets in the activity indicators are disfavored with respect to the model with candidates in RV only.
None of the amplitudes of both signals in activity indicators are significant at 2$\sigma$.

We tried to use an independent phase for activity indicators. The results are in Table \ref{table_signal_activity_indicators_rv}.
We did not find signals in activity indicators at 3$\sigma$, and all the models worsen the lnZ with respect to the model with candidate planets in RVs only. Phase$_b$ calculated in RVs differs from phase$_b$ calculated in $S$ index by 1.25$\sigma$. Phase$_c$ calculated in RVs differs from phase$_c$ calculated in $S$ index by 1.4$\sigma$. Phase$_b$ calculated in RVs differs from phase$_c$ calculated in FWHM by 3.4$\sigma$. Phase$_c$ calculated in RVs differs from phase$_c$ calculated in FWHM by 1.3$\sigma$.
Phase$_b$ calculated in RVs differs from phase$_c$ calculated in $\log{R'_{\rm HK}}$ by 1.8$\sigma$. Phase$_c$ calculated in RVs differs from phase$_c$ calculated in $\log{R'_{\rm HK}}$ by 1.3$\sigma$. Phase$_c$ is poorly defined for both the signals. Phase$_c$ RV = 0.80$_{-0.32}^{+0.28}$ and Phase$_c$ $\log{R'_{\rm HK}}$ = 1.28$_{-0.36}^{+0.14}$.

\begin{table}[]
  \centering
  \begin{threeparttable} 
  \caption{Signals found in activity indicators with a common period to RVs and their significance}
    \label{table_signal_activity_indicators_rv}
    \begin{tabular}{l c c c c}
    \hline
    \hline
    \noalign{\smallskip}
    Indicator &  Phase & $K$$_b$ & $K$$_c$ & $\Delta$lnZ \\
    \noalign{\smallskip}
    \hline
    \noalign{\smallskip}
      $S$ index$\,\times\,1000$ & C & 0.31$^{+0.22}_{-0.19}$ & 0.31$^{+0.22}_{-0.19}$ & $-9$.0 \\ \noalign{\smallskip}
      FWHM\,(\si{\meter\per\second})& C & 0.08$^{+0.11}_{-0.06}$ & 0.14$^{+0.17}_{-0.10}$ & $-8$.8 \\ \noalign{\smallskip}
      $\log{R'_{\rm HK}}$ & C & 1.40$^{+0.92}_{-0.82}$ & 1.42$^{+1.38}_{-0.93}$ & $-1$5.1 \\ \noalign{\bigskip}
      $S$ index$\times$1000 & I & 0.42$\pm$ 0.24 & 0.811$^{+0.39}_{-0.42}$ & $-6$.6 \\ \noalign{\smallskip}
      FWHM\,(\si{\meter\per\second})& I & 0.53$\pm$ 0.21 & 0.47$^{+0.35}_{-0.30}$ & $-4$.1 \\ \noalign{\smallskip}
      $\log{R'_{\rm HK}}$ & I & 2.13$^{+0.95}_{-0.98}$ & 2.71$^{+1.49}_{-1.44}$ & $-1$5.6 \\ \noalign{\smallskip}
      \bottomrule
    \end{tabular}
    \tablefoot{We consider models with phases in common between RVs and activity indicators (C) or independent (I).}
    \medskip
    
    \begin{minipage}{0.5\textwidth}
        \raggedright
        
    \end{minipage}
  \end{threeparttable}
\end{table}
We tried to search for the signal in the activity indicators alone without RVs. We made a multidimensional GP model with two activity indicators. We used uniform priors $\mathcal{U}$(50d,100d) and $\mathcal{U}$(500d,1000d) for signals b and c. We made three models for each activity indicator: only rotation and magnetic cycle, addition of signal b, and addition of signal c. 
We summarize the results in Table \ref{table_signal_activity_indicators_alone}.
\begin{table}[]
  \centering
  \begin{threeparttable} 
  \caption{Amplitude and periods of signals found in the analysis of activity indicators alone.}
    \label{table_signal_activity_indicators_alone}
    \begin{tabular}{l c c c c}
    \hline
    \hline
    \noalign{\smallskip}
    Indicator & K$_b$ & $P_b$ (\si{\day}) & K$_c$ & $P_c$ (\si{\day}) \\
    \noalign{\smallskip}
    \hline
    \noalign{\smallskip}
      $S$ index\,$\times$\,1000 & 0.41 $_{-0.24}^{+0.33}$ & 68 $_{-19}^{+10}$ & 0.81$_{-0.42}^{+0.39}$ & 768$_{-16}^{+33}$ \\ \noalign{\smallskip}
      FWHM  (\si{\meter\per\second})& 0.50$_{-0.31}^{+0.28}$  & 71$_{-9}^{+23}$  & 1.02$_{-0.35}^{+0.40}$ & 765$_{-16}^{+29}$  \\ \noalign{\smallskip}
      $\log{R'_{\rm HK}}$ & 1.73$_{-1.14}^{+1.26}$ & 73$_{-12}^{+17}$  & 3.9$\pm$ 1.4 & 758$_{-31}^{+33}$ \\ \noalign{\smallskip}
      \bottomrule
    \end{tabular}
    \medskip
    
    \begin{minipage}{0.5\textwidth}
        \raggedright
        
    \end{minipage}
  \end{threeparttable}
\end{table}
We searched for the signals in the $S$ index in a multidimensional GP analysis with $S$ index and FWHM. 
In terms of evidence, the one-sinusoidal model is disfavored compared to the stellar activity-only model by $\Delta$lnZ = $-2$.9. The two-sinusoidal model is disfavored compared to the stellar activity-only model by $\Delta$lnZ = $-0$.7. The slight improvement of the two-sinusoidal model over the one-sinusoidal model is due to the 2$\sigma$ detection of a signal at 768$_{-16}^{+33}\,\si{\day}$.

For FWHM, we considered a multidimensional GP with FWHM and $S$ index. 
We found that the one-sinusoidal model is disfavored by $\Delta$lnZ = $-2$.4 and the two-sinusoidal model is disfavored by $\Delta$ lnZ = $-1$.6. 
For $\log{R'_{\rm HK}}$, we used a multidimensional GP model with $\log{R'_{\rm HK}}$ and FWHM. 
We found that the one-sinusoidal model is disfavored by $\Delta$ lnZ = $-2$.6 and the two-sinusoidal model is disfavored by $\Delta$lnZ = $-3$.0.

We plot in Fig. \ref{posterior_planet_b} and Fig. \ref{posterior_planet_c} the posterior distributions of the periods for signals b and c in activity indicators in the two-sinusoidal model, where we see some peaks in the posterior distribution of $P_b$ close to the period of the signal found in RVs.
We show in Table \ref{table_signal_activity_indicators} the significance of the different models we analyzed in activity indicators. 
Signal b is not well defined in period and is not significant in amplitude at 2$\sigma$. Signal c is defined in amplitude at 2$\sigma$ in all the activity indicators we tested, but it points to a different period compared to the signal we found in RVs. 

We do not conclude that the signals in RVs are related to activity. On the other hand, the presence of some hints of signals at the period of candidate b, and the presence of a signal probably related to a harmonic of the cycle at $\sim$ 750\,\si{\day}, makes us cautious. We thus claim two signals we found as candidate planets. We will delegate future work to confirm or reject the planetary hypothesis.

\subsection{Informed search for signals of interest}

\begin{figure}[]
    \centering
    
    \begin{subfigure}[t]{0.48\textwidth}
        \centering
        \begin{tikzpicture}
            \node[anchor=north west, inner sep=0] (image) at (0,0) {\includegraphics[width=\linewidth]{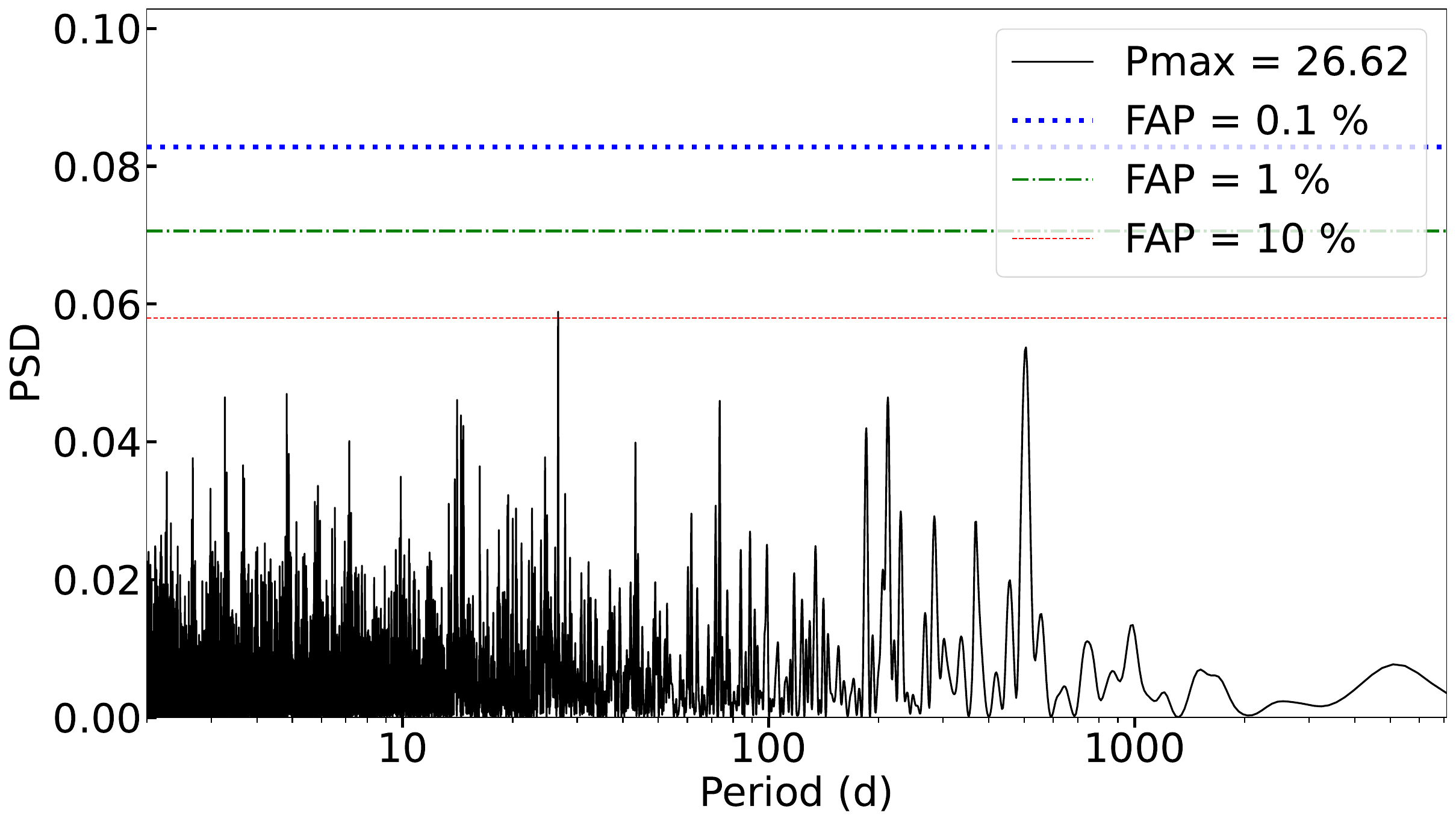}};
            
            \node[anchor=north west, inner sep=0] at (0.15, 0.3) {}; 
        \end{tikzpicture}
    \end{subfigure}
    \caption{GLS periodogram of the RV residuals after subtracting the two-sinusoidal model in an informed search on the period.}
    \label{fig_residuals}
\end{figure}

We made an informed search for the two candidates. We use the parameters found in the current analysis as the final results of our model in Sect. \ref{sec_dis}. We made a multidimensional GP analysis with FWHM and $S$ index as ancillary indicators. We used uniform priors $\mathcal{U}$(50d,100d) and $\mathcal{U}$(500d,1000d) for candidates b and c.
We ran a one-sinusoidal model with the short-period candidate. We found a $\Delta$ lnZ = +9.2 with respect to a flat model. We obtained $K_b$ = 0.67 $\pm$ 0.10\,m\,s$^{-1}$ and $P$ = 72.561 $_{-0.059}^{+0.058}$\,\si{\day}.
We ran a two-sinusoidal model. We found $K_b$ = 0.63 $\pm$ 0.10\,m\,s$^{-1}$ and $P_b$ = 72.578 $_{-0.060}^{+0.059}$\,\si{\day}, $K_c$ = 0.64 $\pm$ 0.12\,m\,s$^{-1}$ and $P_c$ = 682.5$_{-9.9}^{+9.5}$\,\si{\day}. We have a time of inferior conjunction $T_{\rm 0b}$ = 2460877.5$_{-2.9}^{+2.8}$ BJD for the short-period signal and $T_{\rm 0c}$ = 2460236$_{-63}^{+57}$ BJD for the long-period signal. This model was favored compared to the one-sinusoidal model by $\Delta$ lnZ = + 8.6. 
We show in Fig. \ref{fig_residuals} the GLS periodogram of the residuals of the two-sinusoidal model in the case of informed search. We see only a nonsignificant peak at $\sim$ 26\,\si{\day} with false alarm probability (FAP) below 10 \%.
We show the best parameters for the candidates in Table \ref{table_planets}. We show in Table \ref{table_priors} the prior and posterior of the best model adopted. 

We also tried a model with two Keplerians. We followed the parameterization of the eccentricity proposed by \citet{2011_anderson_eccentricity,2013_eastman_eccentricity}, with a combination of e and $\omega$:  $\sqrt{e} \cdot \cos(\omega)$ and $\sqrt{e} \cdot \sin(\omega)$. We used a normal prior $\mathcal{N}$(0,0.3) for the parameters. We imposed the eccentricity to be less than 0.99.
We found the two-Keplerian model to be disfavored by $\Delta$ lnZ = $-1$.4. We found an eccentricity on the short-period Keplerian of 0.19$_{-0.13}^{+0.19}$ and an eccentricity for the long-period Keplerian of 0.41$_{-0.21}^{+0.17}$. Both are not defined at 2$\sigma$.

\begin{table}[]
  \centering
  \begin{threeparttable} 
  \caption{Parameters of the candidate planets of HD 161098.}
    \label{table_planets}
    \begin{tabular}{l c c}
    \hline
    \hline
    \noalign{\smallskip}
    Parameter & HD 161098 b & HD 161098 c \\
    \noalign{\smallskip}
    \hline
    \noalign{\smallskip}
      $T_0$ (BJD)  & 2460877.7$^{+2.8}_{-2.9}$ & 2460236$^{+57}_{-62}$ \\
      $P$ (d) & 72.578$^{+0.059}_{-0.060}$ & 682.5 $_{-9.9}^{+9.5}$ \\
      $K$ (\si{\meter\per\second}) & 0.63 $\pm$ 0.10 & 0.64 $\pm$ 0.12 \\
      $M$$_p$$\sin$i (M$_\oplus$) & 3.63 $\pm$ 0.59 & 7.8$^{+1.5}_{-1.4}$ \\
      $a$ (au) & 0.3207 $\pm$ 0.0037 &  1.429 $\pm$ 0.021 \\
      $S$ ($S_\oplus$) & 6.48$^{+0.42}_{-0.40}$ & 0.326$^{+0.022}_{-0.021}$  \\
      $T_{\rm eq}$ (K) & 406.8 $\pm$ 6.5 & 192.7 $\pm$ 3.2 \\
      \bottomrule
    \end{tabular}
    \tablefoot{The $T_0$ here reported is the time of the inferior conjunction. The temperature is calculated considering an albedo of 0.3.}
    \medskip
    
    \begin{minipage}{0.5\textwidth}
        \raggedright
        
    \end{minipage}
  \end{threeparttable}
\end{table}

\section{Discussion}
\label{sec_dis}
\subsection{Planetary system}
We found two signals of interest at periods of $\sim$ 72.5\si{\day} and $\sim$ 680\si{\day}.
Due to some hints of the presence of the signals in activity indicators, and the doubt on the coherence of the long-period signal in all the datasets, we cannot claim the signals as confirmed planets. We claim the discovery of two candidate planets. Only additional observations will shed light on the nature of the two signals. 

\begin{figure}[]
    \begin{minipage}{0.45\textwidth}
        \includegraphics[width=\linewidth]{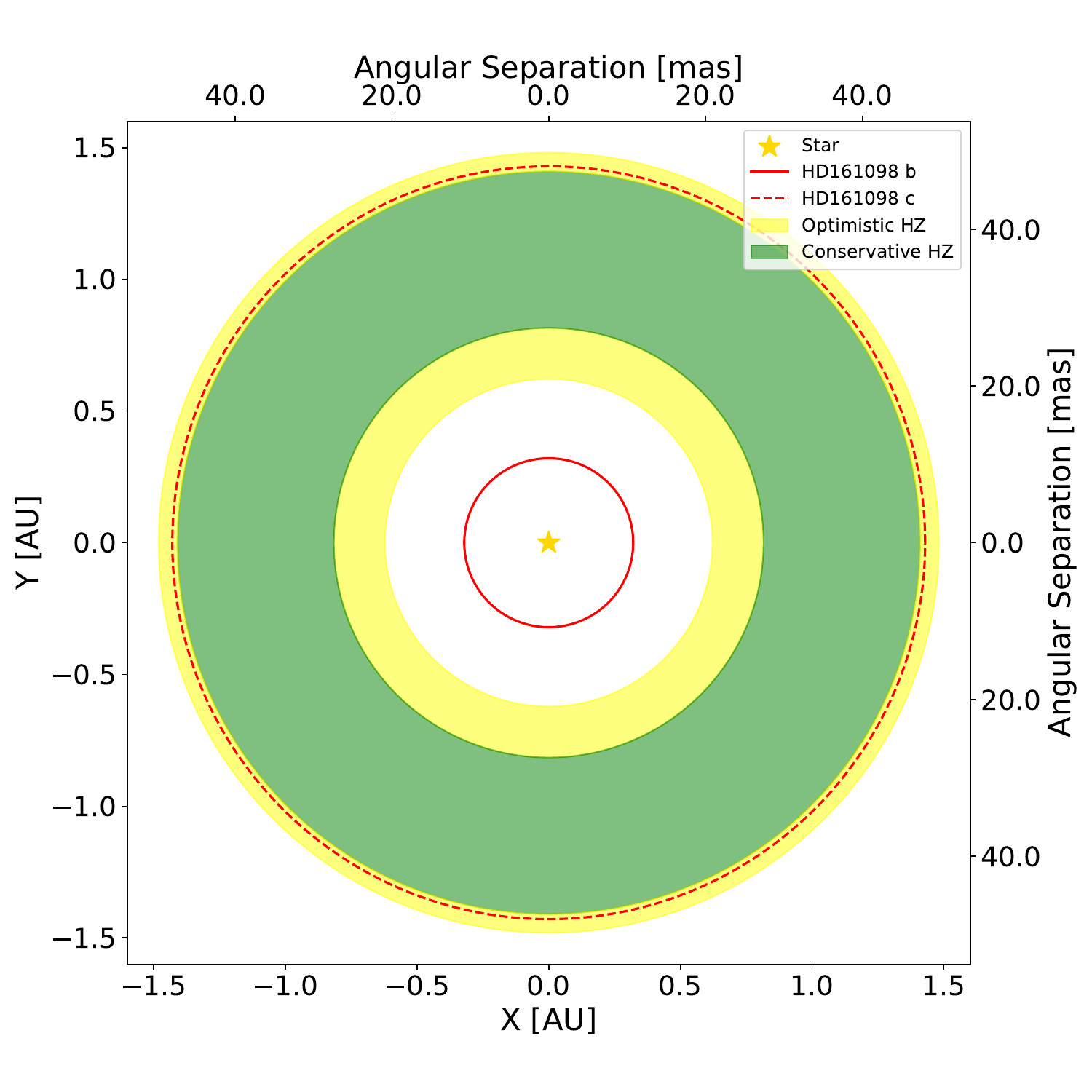}
    \end{minipage}
    \caption{Positions of candidate planets of HD 161098 b and HD 161098 c relative to the HZ of the star.   
    }
    \label{hd161098_planetary_system}
\end{figure}

\begin{figure}[]
    \begin{minipage}{0.45\textwidth}
        \includegraphics[width=\linewidth]{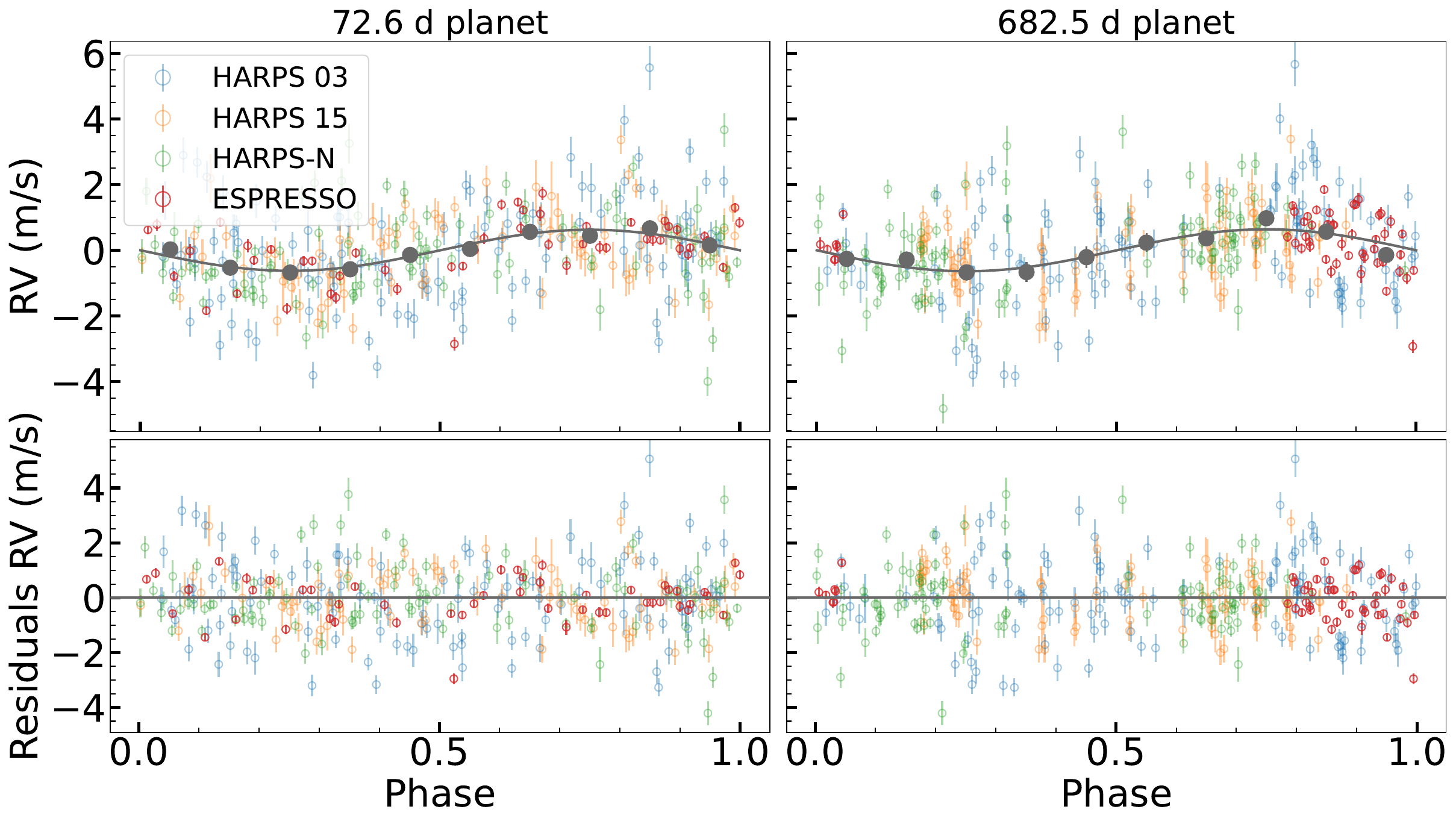}
    \end{minipage}
    \caption{Phase-folded RV plot of candidate planets of HD 161098. 
    }
    \label{phase_folded_fig}
\end{figure}

Candidate HD 161098 b has amplitude $K$$_b$ = 0.63 $\pm$ 0.10\,m\,s$^{-1}$ and period $P_b$ = 72.578$_{-0.060}^{+0.059}$\,\si{\day}. Such a planet would have a minimum mass $M$$_b$ sini = 3.63 $\pm$ 0.59\,$M_\oplus$. The planet would orbit the star at a separation of 0.3209 $\pm$ 0.003\, au. 
It would receive an insolation of 6.48$_{-0.40}^{+0.42}$\,$S_\oplus$ and it would have an equilibrium temperature of 406.8 $\pm$ 6.5\,K following Eq. \ref{temp_equation} \citep{2010_seager_temp}:
\begin{equation}
T_{\mathrm{eq}} = T_{\star}
\sqrt{\frac{R_{\star}}{2a}}
\left(1 - A\right)^{1/4}
\label{temp_equation}
.\end{equation}
\noindent
We considered a Bond albedo of 0.3. 
Candidate HD 161098 c has an amplitude $K$$_c$ = 0.64$\pm$ 0.12\,m\,s$^{-1}$ and an orbital period $P_c$ = 682.5$_{-9.9}^{+9.5}$\,\si{\day}. If it were a planet it would have a minimum mass of $M$$_c$sini = 7.8$_{-1.4}^{+1.5}$\,$M$$_\oplus$. The planet would orbit the star at a separation of 1.429 $\pm$ 0.021\,au.
It would receive an insolation of 0.326$_{-0.021}^{+0.022}$\,$S$$_\oplus$ and it would have an equilibrium temperature of 192.7 $\pm$ 3.2\,K. HD 161098 c would reside between the boundaries of the conservative and optimistic outer habitable zone. We show in Fig. \ref{hd161098_planetary_system} the position of the planets relative to the habitable zone. 
We measured $v \sin{i}$ = 1.0 $_{-0.4}^{+0.3}$ \,km\,s$^{-1}$. 
To measure vsini, we used the FWHM of the CCF in a way similar to that done in \citet{2023_rainer_vsini}.
We determined the inclination of the star with the method described by \citet{masuda_2020_inclination}. We found an angle $I$ = 51 $\pm$ 20\,deg between the line of sight and the rotation axis of the star. With coplanar planets we would have true masses M$_b$ = 4.7$_{-1.1}^{+2.4}$\,M$_\oplus$ and M$_c$ = 10.2$_{-2.5}^{+5.3}$\,M$_\oplus$.
Candidate b would be a super-Earth, with different possible compositions. Candidate c would be a sub-Neptune-like planet.
Fig. \ref{phase_folded_fig} shows the phase-folded plot of the two candidates. Fig. \ref{phase_folded_fig_zoom} shows a zoomed-in image of the same plot. 
We show in Table \ref{residuals_rms_table} the evolution of the root mean square (RMS) of the residuals of RVs after subtracting the different contributions: 
activity-related terms, HD 161098 b, and HD 161098 c. The RMS of the residuals always decreases for each dataset at every step.
In Table \ref{table_planets} we show the parameters of the two candidate planets. 
In Table \ref{table_priors} we show the priors used in our analysis with the results of the analysis. The candidate HD 161098 c would reside in the optimistic HZ of its parent star. Few discovered planets reside in the HZ of stars hotter than 4000\,K. We show in Fig. \ref{fig_literature_hd161098} the plot of all the planets with masses below 20 M$_\oplus$ and insolation between 0.1\,S$_\oplus$ and 10\,S$_\oplus$ orbiting around stars hotter than 4000\,K. We used this limit to circumscribe the research to G- and K-type stars.

\subsection{Detection limits}
\label{det_lim_sect}
\begin{figure}[]
    \centering
    {\includegraphics[width=\linewidth]{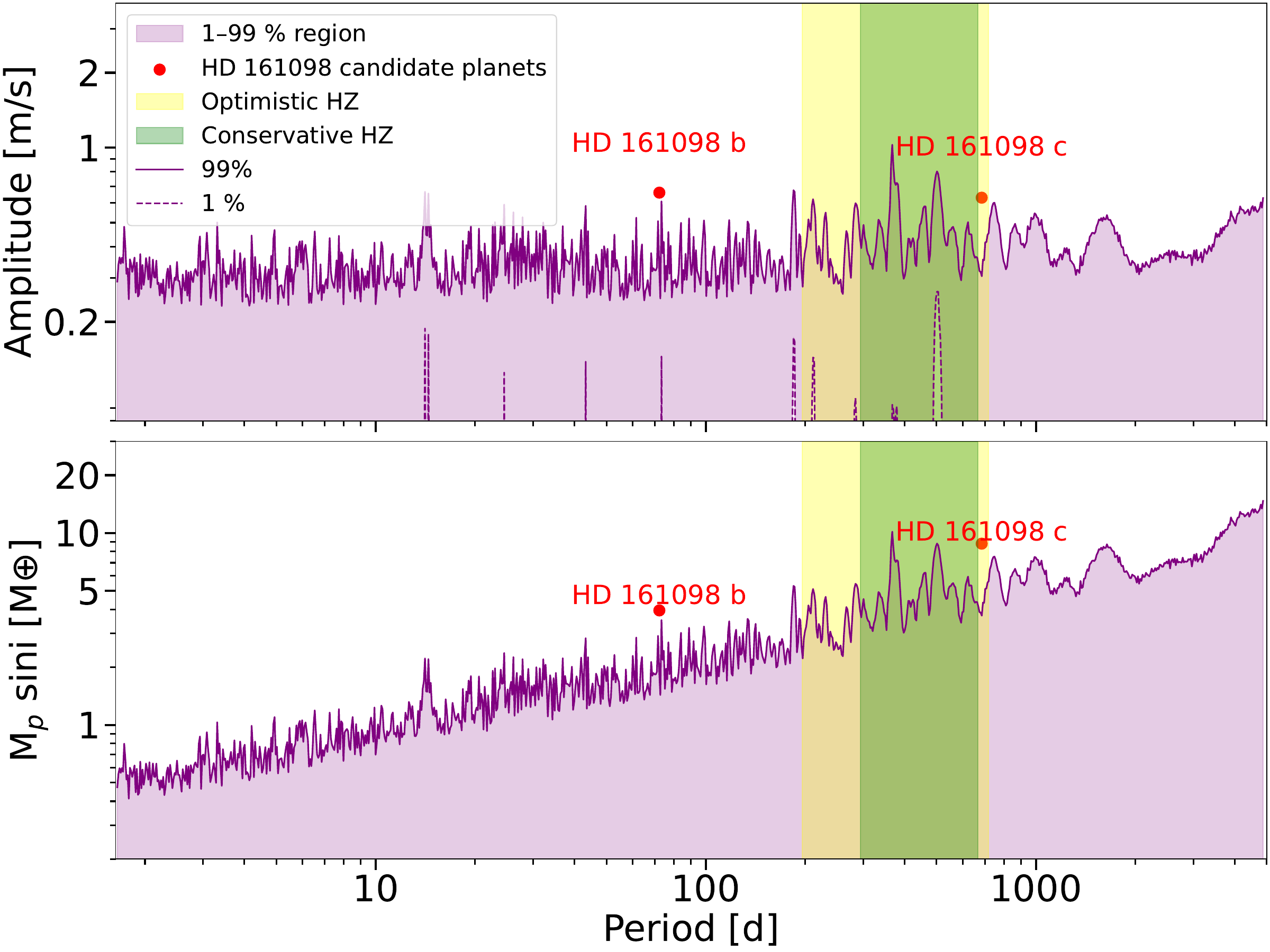}}
    \hfill
    \caption{Detection limits for HD 161098. Top panel: Detection limits in amplitude. 
    Bottom panel: Detection limits in mass.
    }
    \label{detection_limits_fig}
\end{figure}

We followed \citet{2026_nari_hd176986} to measure the detection limits. We took a grid of 1000 periods uniformly sampled in logarithmic space between $\sim$ 1.5\,\si{\day} and $\sim$ 5000\,\si{\day}. We fixed the period of the sinusoidal from the value taken from the grid, and we considered the 99th percentile of the posterior distribution of the amplitude as our detection limit. We show in Fig. \ref{detection_limits_fig} the results of our analysis. 
We found a mean detection limit in amplitude of $\sim$ 38 \si{\centi\meter\per\second}. At a few periods, the 1st percentile of the posterior of the amplitude is above 10 \si{\centi\meter\per\second}. These outliers could be related to additional noise not fitted by our model, but could also hide the presence of low-amplitude signals of a different origin. In mass, we can detect planets with minimum mass inferior to 1 $M$$_\oplus$ up to almost 10\,\si{\day}, while our sensitivity to planets with minimum mass below 10\,$M$$_\oplus$ spans until $\sim$ 2000\,\si{\day}. We can detect planets with masses below 20 $M$$_\oplus$ up to 5000\,\si{\day} orbital period. 

\subsection{Stellar activity}

In the multidimensional GP analysis we found $P_{\rm cycle}$ = 4090$_{-130}^{+140}$\,\si{\day} and $P_{\rm rot}$ = 28.22$_{-0.35}^{+0.30}$\,\si{\day}. 
The magnetic cycle is longer than one-half the timespan of the observations. Future observations will refine the $P_{\rm cycle}$.
We found a timescale of evolution equal to 107$_{-21}^{+26}$\,\si{\day}. This is $\sim$ 3.8 times the rotation period of the star. 
This value is different from that of the Sun, and permits us to classify HD 161098 as a beater star, following the classification of \citet{2017_giles_beater}. A Sun-like star has a timescale of decay of the spots of approximately one rotation, while for a beater we have a lifetime of spots of a few rotations. A long decay time of the spots can suggest the presence of polar spots on the surface of the star \citep{2017_giles_beater}.

We analyzed the dispersion in the residuals of the activity indicators and RVs at different phases of the cycle. First, we considered the dispersion between BJD 2455500 and BJD 2456500, at the maximum of the activity of the cycle, and then between BJD 2458000 and BJD 2459500, in a phase of a minimum of activity of the cycle. At the minimum of the cycle region, we have 41 epochs. At the maximum of the cycle, we have 59 epochs of observation. We derived these two regions from visual inspection of Fig. \ref{hd161098_activity_indicators}. 
We used the RMS of the residuals as our metric of reference. We report the results of the analysis in Table \ref{table_min_max}.
In RVs we see a decrease in the dispersion of the residuals at the minimum of the cycle, while we see an increase in the dispersion for both FWHM and $S$ index. We need to take into account multiple factors. The two regions do not have observations from overlapping instruments. The H15 dataset shows a larger dispersion in the $S$ index compared to H03. Both the regions of minimum and maximum of the cycle show a standard deviation of the residuals comparable to or larger than the standard deviation of the full dataset, except for the $S$ index at the maximum of the cycle. 
In the raw RVs, we see a lower dispersion at the minimum of the cycle, but this is not strongly confirmed in the activity indicators. We cannot conclude that observing a star at the minimum of its magnetic activity is related to a lower dispersion of the dataset. We cannot exclude this hypothesis, due to external factors such as the use of different instruments and the nonhomogeneity of the time series.

\begin{table}[]
  \centering
  \begin{threeparttable} 
  \caption{RV and activity indicator dispersion at the minimum and maximum of the cycle.}
    \label{table_min_max}
    \begin{tabular}{l c c c }
    \hline
    \hline
    \noalign{\smallskip}
    Dataset & Min cycle  & Max cycle & Full \\
    \noalign{\smallskip}
    \hline
    \noalign{\smallskip}
      RV RMS\,(\si{\meter\per\second}) & 1.51 & 2.21 & 1.72 \\ 
      RV residuals RMS\,(\si{\meter\per\second}) & 1.42
      & 1.69 & 1.21 \\ 
      $S$ index RMS & 4.41 & 4.18 & 5.88 \\
      $S$ index residuals RMS & 3.21 & 1.89 & 2.42\\
      FWHM RMS\,(\si{\meter\per\second}) & 3.48 & 3.84 & 4.28 \\ 
      FWHM residuals RMS\,(\si{\meter\per\second}) & 2.66 & 2.41 & 2.59 \\ 
      \bottomrule
    \end{tabular}
    \medskip
    
    \begin{minipage}{0.5\textwidth}
        \raggedright
        
    \end{minipage}
  \end{threeparttable}
\end{table}

\section{Conclusion}

We made an in-depth analysis of the star HD 161098. We found evidence of two candidate planets. HD 161098 b has a period of 72.578$_{-0.060}^{+0.059}$\,\si{\day}, an amplitude of 0.63 $\pm$ 0.10\,m\,s$^{-1}$, a minimum mass of 3.63$\pm$ 0.59\,M$_\oplus$; it orbits the star at a separation of 0.3207$\pm$ 0.0037 au. HD 161098 c has a period of 682.5$_{-9.9}^{+9.5}$\,\si{\day}, an amplitude of 0.64 $\pm$ 0.12\,m\,s$^{-1}$, a minimum mass of 7.8$_{-1.4}^{+1.5}$\,M$_\oplus$, and orbits the star at a separation of 1.429 $\pm$ 0.021 au. We found the inclination of the system to be I = 51$\pm$ 20\,\si{\degree}. This implies, in the case that the two candidates are confirmed, that we are in the presence of a super-Earth in the inner orbit and a super-Earth or a mini-Neptune planet in the outer orbit. The outer candidate resides slightly outside the conservative HZ of the star, in the optimistic HZ. 

HD 161098 b is stable through the observing seasons. For HD 161098 c, we see a preference for the H03 dataset. We tried to recover the signal without H03, and we could only find the signal at the correct period if we considered a normal prior on the period. The results of the apodized test do not lead us to claim the signal as a confirmed detection. 
We searched for the candidate signals in the activity indicators. For both HD 161098 b and HD 161098 c, we found the signal to be at $\sim$ 2$\sigma$ level in the activity indicators only when we used a different phase between the RV signal and the signal in the activity indicators. Due to this test, we point out the need for additional observations to confirm the planetary nature of the two signals.

We measured the detection limits of the dataset. We found a detection limit below 1\,$M$$_\oplus$ up to an orbital period of $\sim$ 10\,\si{\day}. We can detect of planets with minimum mass below 20\,$M$$_\oplus$ up to 5000\,\si{\day}. 

We checked the differences in the dispersion of the residuals of RVs and activity indicators between the maximum and the minimum of the magnetic cycle. We did not find significant evidence that the observations collected at the minimum of the cycle are better suited for the detection of exoplanets.
\section{Data availability}
A table for RVs and activity indicators is only available in electronic form at the CDS via anonymous ftp to cdsarc.u-strasbg.fr (130.79.128.5) or via http://cdsweb.u-strasbg.fr/cgi-bin/qcat?J/A+A/
\label{sec_con}

\begin{acknowledgements}
      NN acknowledges funding from Light Bridges for the Doctoral Thesis "Habitable Earth-like planets with ESPRESSO and NIRPS", in cooperation with the Instituto de Astrofísica de Canarias, and the use of Indefeasible Computer Rights (ICR) being commissioned at the ASTRO POC project in the Island of Tenerife, Canary Islands (Spain). The ICR-ASTRONOMY used for his research was provided by Light Bridges in cooperation with Hewlett Packard Enterprise (HPE). 
      JIGH, AKS, RR, CAP, NN, VMP, and ASM acknowledge financial support from the Spanish Ministry of Science and Innovation (MICINN) project PID2020-117493GB-I00. The project that gave rise to these results received the support of a fellowship from the ”la Caixa” Foundation (ID 100010434). The fellowship code is LCF/BQ/DI23/11990071.
      This publication makes use of The Data \& Analysis Center for Exoplanets (DACE), which is a facility based at the University of Geneva (CH) dedicated to extrasolar planets data visualization, exchange, and analysis. DACE is a platform of the Swiss National Centre of Competence in Research (NCCR) PlanetS, federating the Swiss expertise in Exoplanet research. The DACE platform is available at \url{https://dace.unige.ch}.
      XD acknowledges the support from the European Research Council (ERC) under the European Union’s Horizon 2020 research and innovation programme (grant agreement SCORE No 851555) and from the Swiss National Science Foundation under the grant SPECTRE (No $200021\_215200$). This work has been carried out within the framework of the NCCR PlanetS supported by the Swiss National Science Foundation under grants $51NF40\_182901$ and $51NF40\_205606$.
      This research has extensively used the SIMBAD database operated at CDS, Strasbourg, France,
      and NASA’s Astrophysics Data System. This research has made use of NASA
      Exoplanet Archive, which is operated by the California Institute of Technology, under contract with the National Aeronautics and Space Administration
      under the Exoplanet Exploration Program. The manuscript was written using \texttt{Overleaf}. Extensive use of \texttt{numpy} \citep{2011_numpy} and \texttt{scipy} \citep{2020_virtanen_scipy}.
      The main analysis was performed in Python 3 \citep{python3_ref} running on a Ubuntu system \citep{ubuntu_2015}.
      Programs IDs for the HARPS observations we used in the analysis are: 072.C-0488(E), 091.C-0936(A), 183.C-0972(A), 192.C-0852(A), Udry, 105.20PH.001, 109.2392.001, 112.25SF.001, 113.26U2.001, and 198.C-0836(A). Programs IDs for the HARPS-N observations we used in the analysis are: CAT14A\_83, CAT15A\_140, CAT16A\_109, CAT17A\_38, CAT17A\_58, CAT18A\_115, CAT19A\_159, CAT20A\_121, and ITP15\_7.
      ESPRESSO observations we used in the analysis have program ID 115.28BD.
\end{acknowledgements}

\bibliographystyle{aa} 
\bibliography{bibliography_hd161098}

\begin{appendix}
\section{Methods}
\label{methods_sec}
For the parameter estimation, we used the nested-sampling tool {\tt Dynesty} \citep{2020_destiny_speagle}. {\tt Dynesty} gives an estimate of the natural logarithm of the evidence associated with a model, allowing an easy model comparison. We have considered as a criterion to accept a more complex model an improvement in lnZ of 5 or more. We use for the inference of parameters a number of livepoints equal to the maximum between 500 and 40 times the number of parameters of the model. For the FIP analysis of Sect. \ref{fip_sect}, we have used a number of live points equal to 5 times the difference between the maximum and minimum frequencies we consider for the model, divided by the frequency resolution measured as $2 \pi$/(max(RV)--min(RV)). In this way, we have 5 live points per element of frequency resolution. We have used it as a stopping criterion for convergence when the sampler remains with less than $\Delta$ lnZ = 0.01 to explore. To account for the different zero-points of different instruments, we always consider an offset term for each instrument in the analysis. We consider an offset specifically for the FWHM for the HN dataset after BJD 2459500. This is due to an inspection of the behavior of the FWHM of HN for multiple stars.
A jitter term is added in quadrature to the nominal error of the different instruments for each time series. The jitter term takes into account all the sources of noise we are not modeling for, and the instrumental noise. To remove outliers from different datasets, we bin the observations nightly and we apply a cut on the dataset consisting of a 5$\sigma$ clipping joined with the exclusion of measurements where the error is larger than three times the median error for each dataset. To implement Gaussian processes in our analysis we used {\tt S+LEAF} \citep{2022_spleaf_delisle}. {\tt S+LEAF} allows only a certain kind of semi-separable matrix to be taken into account as a covariance matrix. In this way, the computational cost scales linearly with the dimension of the dataset, instead of with its cube, as it used to be in standard implementations. Before a fit for the jitter was available, exploratory GLS periodograms of our datasets were generated adding in quadrature a white noise term to the error on the measurements equal to the standard deviation of the dataset. 
For parameter determination we considered the median of the posterior as the final value of the parameter and the 16th and 84th percentile of the posterior as lower and upper error.
We have used the package for the GLS periodogram described by \citet{2009_gls_zeichmeister}. We split the FWHM HN dataset, applying a different zero-point for observations taken before 2459500 BJD and observations taken after this date. This is due to an offset present only in the FWHM of HARPS-N.

\section{Stellar activity}
\label{stellar_activity_appendix}

Extremely precise spectrographs at 1\,m\,s$^{-1}$ precision made effects from the star become detectable in the RV time series. Stellar systems are nonequilibrium systems on a variety of timescales. 
\\
Stellar oscillations can originate short-term RV variations on the timescale of minutes with amplitudes up to tens of \si{\centi\meter\per\second} \citep{2008_otoole_oscillations}.
We can use tailored observing times to mitigate the stellar pulsation effect \citep{2011_dumusque_oscillations_granulations,2019_chaplin_oscillations}. 

Granulation generates RV variations with a timescale of a few minutes to multiple days and amplitudes of a few m\,s$^{-1}$ \citep{2011_dumusque_oscillations_granulations,2011_mathur_granulation}. 
Multiple observations of the same target through the same night can partially average out the effect of granulation
\citep{2011_dumusque_oscillations_granulations}. In large surveys, due to time constraints, is not always possible to implement this method, and the granulation effect remains a barrier toward the detection of tiny signals.

The Sun hosts an 11yr magnetic cycle \citep{1844_schwabe_solar_cycle}. \citet{2011_lovis_cycle} shows this is common in other stars too. The amplitude of magnetic cycles signals in RV time series can reach tens of m\,s$^{-1}$ and mimic long-period planets. We modeled the cycle component in our time series with a sinusoidal and, eventually, additional harmonics.  

The rotation of the star is an important source of stellar noise in RV time series. Inhomogeneities in stellar flux and suppression of convection associated with starspots create stellar-related RV variations \citep{1997_saar_activity}. The amplitude of the effect of stellar rotation on RV can reach tens of m\,s$^{-1}$ \citep{1997_saar_activity}. These variations can be quasi-periodic and mimic or hide planetary signals \citep{2021_meunier_activity}.  

We use Gaussian processes (GP) to model the rotation signal in RVs \citep{2006_rasmussen_gp_book,2023_aigrain_gp}.
Instead of parametrizing an analytical function to interpolate the signal of activity, GPs sample for the covariance between observations $\boldsymbol{K}$ = k($t_i$,$t_j$,$\phi$), where k($t_i$,$t_j$,$\phi$) is the covariance between observations made at time $t_i$ and $t_j$, and $\phi$ are the hyper-parameters of the covariance function. The covariance function is the kernel of the GP.

A common covariance function used in the field of exoplanets to model rotation-related activity is the quasi-periodic kernel defined as:
\begin{equation}
k(t, t') = \sigma^2 \exp\left(-\frac{(t - t')^2}{2\lambda^2}\right) \exp\left(-\frac{\sin^2\left(\frac{\pi |t - t'|}{P}\right)}{2\Gamma^2}\right)
\end{equation}
\noindent
Where $\sigma$ is the amplitude of the kernel, $\lambda$ is the timescale of evolution of the correlations, $P$ is the rotation period, and $\Gamma$ is the harmonic complexity. We used the MEP kernel, an approximation of the quasi-periodic kernel. The MEP kernel is representable as an S+LEAF matrix, so a matrix which is a sum of semi-separable and a LEAF matrix \citep{2020_spleaf_delisle}. This decreases the computational cost from $\mathcal{O}(n^3)$ to $\mathcal{O}(n)$, with n the number of measurements.

To mitigate the tendency of the GPs to overfit \citep{2026_nari_hd176986}, we can train them together with activity indicators in a multidimensional GP fashion \citep{2015_rajpaul_multigp,2023_barragan_multigp}. This framework was first inspired by the FF' method \citep{2012_aigrain_ff'}. The \textit{FF'} method models the RV variation with a linear combination of the squared flux obtained in simultaneous photometry and the flux multiplied by its time derivative: 
\begin{equation}
\text{RV} = V_r F(t) \dot{F}(t) + V_c F^2(t)
\end{equation}
\noindent
The term with the derivative represents the flux modulation coming from the break of the flux symmetry and the term with $F^2(t)$ is linked to the suppression of convective blueshift in magnetized regions.

In the multidimensional GP framework, both RV and activity indicators are modeled with a linear combination of an underlying GP and its first derivative:
\begin{align}
\mathrm{RV} &= V_c G(t) + V_r \dot{G}(t), \\
A_1 &= A_{1c} G(t), \\
A_2 &= A_{2c} G(t) + A_{2r} \dot{G}(t).
\end{align}
\noindent
where V$_c$, V$_r$, A$_{1c}$, A$_{2c}$, and A$_{2r}$ are the coefficients of RV and two hypothetical activity indicators. 

\begin{table}[h!]
  \caption[]{Stellar parameters of interest for HD 161098.}
  \label{tab_stel_par}
  \begin{tabular}{p{0.5\linewidth}ll}
    \hline
    \hline
    \noalign{\smallskip}
    Parameter & HD 161098 & Ref\\
    \noalign{\smallskip}
    \hline
    \noalign{\smallskip}
    $\alpha$ & 17:44:25.252 $\pm$ 0.019 & 1\\ 
    $\delta$ & $-0$3:55:04.315 $\pm$ 0.015 & 1\\
    Parallax\,(mas) & 33.61 $\pm$ 0.02 & 2 \\ 
    $d$\,(pc) & 29.750 $\pm$ 0.018 & 1 \\
    $\mu_{\rm \alpha}$ $\cos$ $\delta$ (mas yr$^{-1}$) & $-1$08.503 $\pm$ 0.023 & 1 \\
    $\mu_{\delta}$ (mas yr$^{-1}$) &  $-2$40.777 $\pm$ 0.017 & 1 \\
    $T_{\rm eff}$\,(K) & $5610 \pm 50$ & 3 \\
    $\log_{10} g$\,(cgs) & $4.460 \pm 0.083$ & 3 \\ 
    Spectral type & G8\,V & 4\\
    $\gamma$\,(\si{\kilo\meter\per\second}) & $-1$6.9940 $\pm$ 0.0004 & 5\\ 
    $[$Fe/H$]$ (dex) & $-0.240 \pm 0.049$ & 3\\
    $M_{\star}$ (M$_{\odot}$) & 0.837 $\pm$ 0.029 & 6\\ 
    $R_{\star}$ (R$_{\odot}$) & 0.866$_{-0.020}^{+0.022}$ & 7\\
    $\log{R'_{\rm HK}}$ & $-4$.894 $\pm$ 0.011 & 0 \\
    $B$\,(mag) & 8.34 $\pm$ 0.02 & 8 \\
    $V$\,(mag) & 7.67 $\pm$ 0.01 & 8 \\
    $L_{\rm bol}$\,(L$_{\odot}$) & 0.6769 $\pm$ 0.0015 & 7 \\ 
    Age\,(Gyr) & 8.3 $\pm$ 3.9 & 6 \\
    P$_{\rm rot}$\,(d) & 28.22$_{-0.35}^{+0.30}$ & 0 \\
    P$_{\rm cycle}$\,(d) & 4090$_{-130}^{+140}$ & 0 \\
    Inner optimistic HZ\,(au) &  0.6216 $\pm$ 0.0014 & 0 \\
    Inner conservative HZ\,(au) & 0.8155 $\pm$ 0.0025 & 0 \\
    Outer conservative HZ\,(au) & 1.4097$_{-0.0062}^{+0.0063}$ & 0 \\
    Outer optimistic HZ\,(au) & 1.4808 $\pm$ 0.0066 & 0 \\
    \noalign{\smallskip}
    \hline
  \end{tabular}
  \medskip 
    \begin{minipage}{0.5\textwidth}
        \raggedright
        References: 0 - This work, 1 - \citealp{2020_gaia_edr3}, 2 -\citealp{2023_vallenari_gaia_dr3}, 3 - \citealp{2024_perdelwitz_temp}, 4 - \citealp{1999_michigan_catalogue}, 5 - \citealp{2018_gaia_rv}, 6 -
        \citealp{2019_delgado_mena_mass}, 7 -
        \citealp{2018_gaia_dr2}, 8 -
        \citealp{2000_tycho_catalogue}.
    \end{minipage}
\end{table}
\section{Figures and tables}

\begin{figure*}[]
    \begin{minipage}{\textwidth}
        \includegraphics[width=0.9\linewidth]{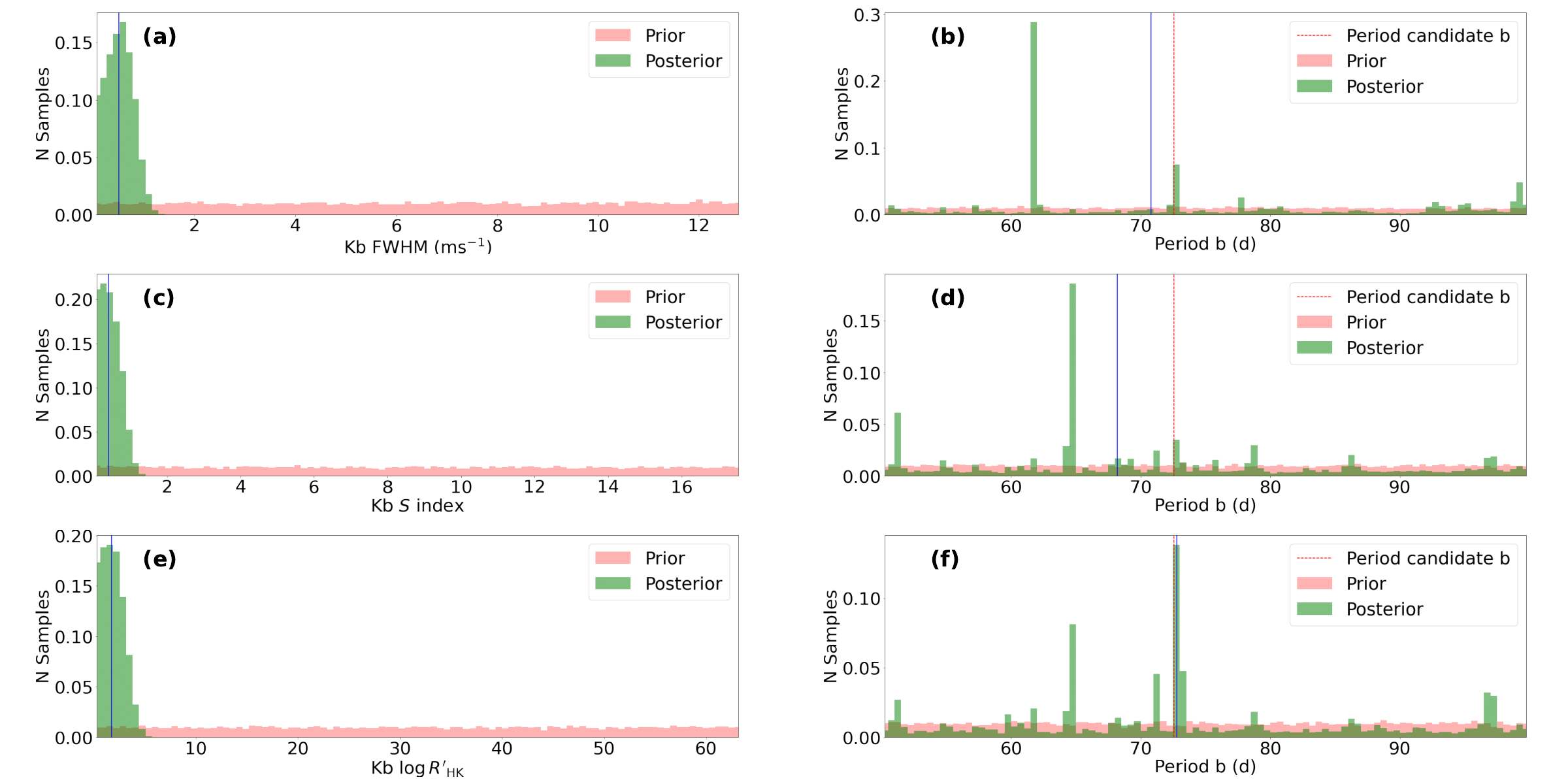}
    \end{minipage}
    \caption{Search for candidate planets signals in activity indicators, posterior distribution. The blue vertical lines represent the median value of the posterior distribution. Panel a) Posterior distribution of candidate signal b in FWHM. Panel b) Posterior distribution of the period of candidate b in FWHM. Panel c) Posterior distribution of the amplitude of candidate b in $S$ index. Panel d) Posterior distribution of the period of the candidate b in $S$ index. Panel e) Posterior distribution of the amplitude of candidate b in $\log{R'_{\rm HK}}$. Panel f) Posterior distribution of the period of candidate b in $\log{R'_{\rm HK}}$.}
\label{posterior_planet_b}
\end{figure*}

\begin{figure*}[]
    {\includegraphics[width=0.9\linewidth]{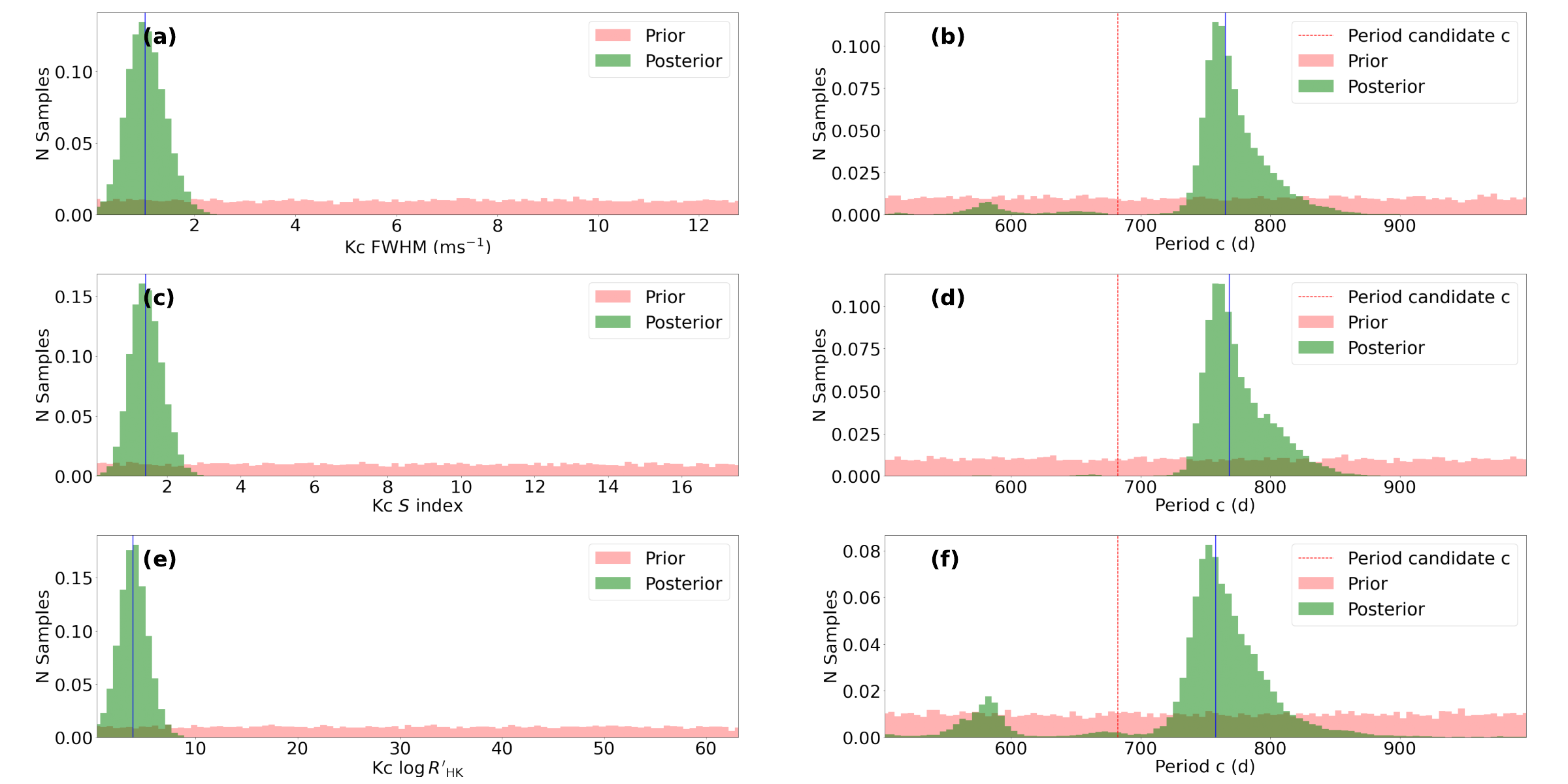}}
    \hfill
    \caption{Search for candidate planet signals in activity indicators, posterior distribution. The blue vertical lines represent the median value of the posterior distribution. Figure a) Posterior distribution of candidate signal c in FWHM. Panel b) Posterior distribution of the period of candidate c in FWHM. Panel c) Posterior distribution of the amplitude of candidate c in $S$ index. Panel d) Posterior distribution of the period of the candidate c in $S$ index. Panel e) Posterior distribution of the amplitude of candidate c in $\log{R'_{\rm HK}}$. Panel f) Posterior distribution of the period of candidate c in $\log{R'_{\rm HK}}$.}

\label{posterior_planet_c}
\end{figure*}

\begin{figure}[]
    \begin{minipage}{0.45\textwidth}
        \includegraphics[width=\linewidth]{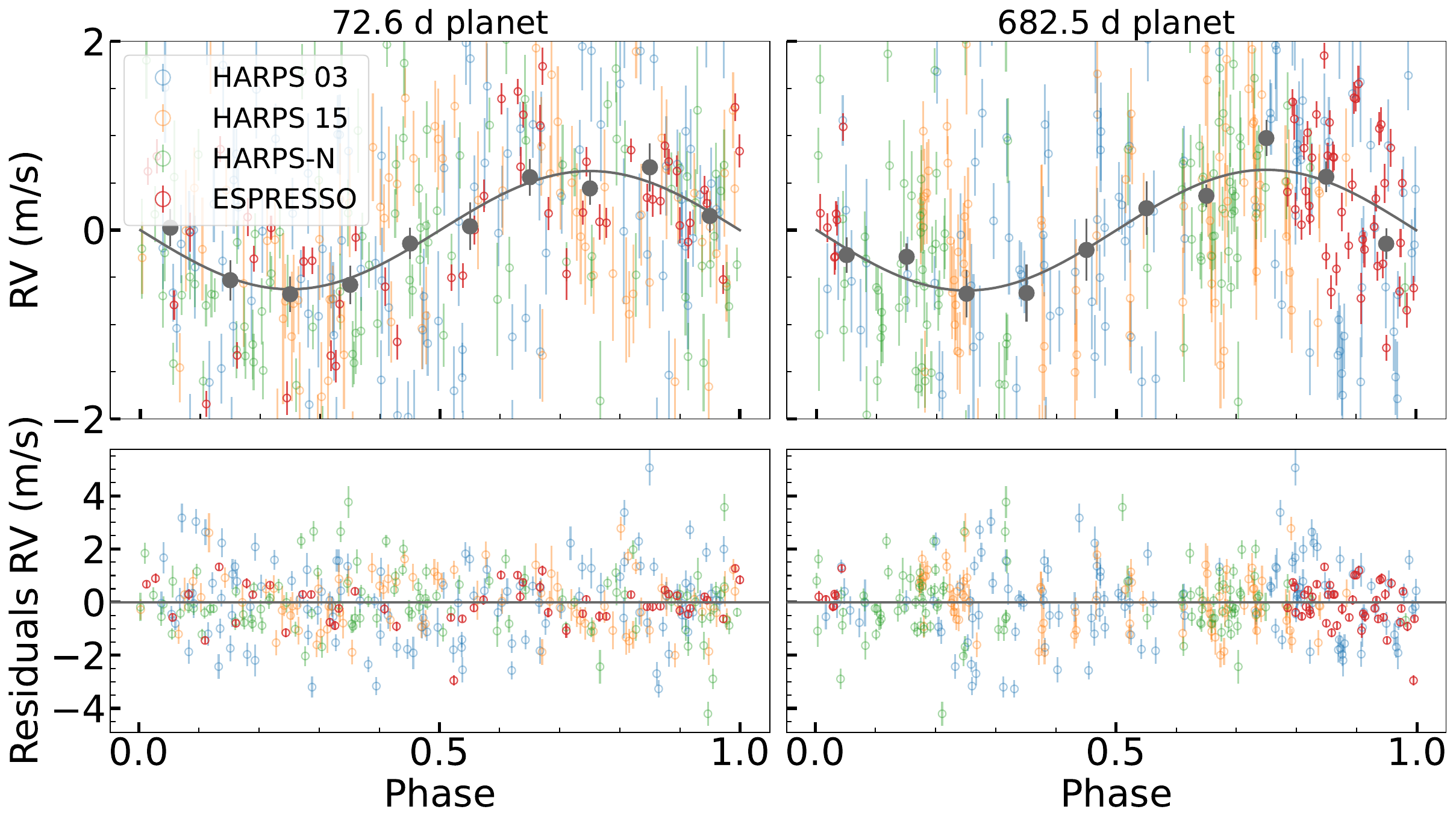}
    \end{minipage}
    \caption{Zoomed-in image of the phase-folded plot of candidate planets of HD 161098. 
    }
    \label{phase_folded_fig_zoom}
\end{figure}

\begin{table}[]
  \centering
  \begin{threeparttable} 
  \caption{Significance of stellar activity-only models when searching for signals with periods compatible with the periods of the candidates.}
    \label{table_signal_activity_indicators}
    \begin{tabular}{l r}
    \hline
    \hline
    \noalign{\smallskip}
    Model  &  $\Delta$lnZ \\
    \noalign{\smallskip}
    \hline
    \noalign{\smallskip}
      $S$ index stellar activity  & 0 \\
      $S$ index stellar activity + one sinusoidal & $-2$.9 \\
      $S$ index stellar activity + two sinusoidals & $-0$.7 \\
      & \\
      FWHM stellar activity  & 0 \\
      FWHM stellar activity + one sinusoidal & $-2$.4 \\
      FWHM stellar activity + two sinusoidals & $-1$.6 \\
      & \\
      $\log{R'_{\rm HK}}$ stellar activity  & 0 \\
      $\log{R'_{\rm HK}}$ stellar activity + one sinusoidal & $-2$.6 \\
      $\log{R'_{\rm HK}}$ stellar activity + two sinusoidals & $-3$.0 \\
      \bottomrule
    \end{tabular}
    \medskip
    
    \begin{minipage}{0.5\textwidth}
        \raggedright
        
    \end{minipage}
  \end{threeparttable}
\end{table}

\begin{table*}[]
  \centering
  \begin{threeparttable}
    \caption{Priors and posteriors of our favorite model.} 
    \label{table_priors}
    \small
    \begin{tabular}{l r r r}
    \hline
    \hline
    \noalign{\smallskip}
    Parameter & Unit & Prior & Posterior \\
    \noalign{\smallskip}
    \hline
    \noalign{\smallskip}
      K$\rm_b$  & m\,s$^{-1}$ & $\mathcal{U}(0,3\sigma RV)$ & 0.63 $\pm$ 0.10 \\
      $P\rm_b$ & d & $\mathcal{U}(50,100)$ & 72.578$^{+0.059}_{-0.060}$ \\
      Phase $_b$ &  & $\mathcal{U}(0.5,1.5)$ & 0.992$_{-0.039}^{+0.041}$ \\
      K$\rm_c$  & m\,s$^{-1}$ & $\mathcal{U}(0,3\sigma RV)$ & 0.64 $\pm$ 0.12 \\
      $P\rm_c$ & d & $\mathcal{U}(500,1000)$ & 682.5 $_{-9.9}^{+9.5}$ \\
      Phase $\rm_c$ & & $\mathcal{U}(0.5,1.5)$ & 1.04$_{-0.09}^{+0.11}$ \\
      K1$\rm_{cycle}$ RV & m\,s$^{-1}$ & $\mathcal{U}(0,3\sigma RV)$ & 0.87$_{-0.30}^{+0.28}$ \\
      $P\rm_{cycle}$ & \si{\day} & $\mathcal{U}(2000,6000)$ &  4090$_{-130}^{+140}$\\
      Phase 1 $\rm_{cycle}$ & & $\mathcal{U}(0.5,1.5)$ & 0.875$_{-0.040}^{+0.039}$\\
      K2$\rm_{cycle}$ RV & m\,s$^{-1}$ & $\mathcal{U}(0,3\sigma RV)$ & 0.46$_{-0.21}^{+0.25}$ \\
      Phase 2 $\rm_{cycle}$ & & $\mathcal{U}(0.0,1.0)$ & 0.282$_{-0.080}^{+0.091}$\\
      RV offset H03 & m\,s$^{-1}$ & $\mathcal{N}(0,3\sigma RV)$ & 0.32 $_{-0.25}^{+0.26}$ \\
      RV offset H15 & m\,s$^{-1}$ & $\mathcal{N}(0,3\sigma RV)$ & $-0$.18 $\pm$ 0.32 \\
      RV offset HN & m\,s$^{-1}$ & $\mathcal{N}(0,3\sigma RV)$ & 0.53 $_{-0.26}^{+0.28}$ \\
      RV offset E19 & m\,s$^{-1}$ & $\mathcal{N}(0,3\sigma RV)$ & $-0$.64 $\pm$ 0.72 \\
      RV log jitter H03 & & $\mathcal{U}(-5,5)$& 0.399$_{-0.081}^{+0.082}$\\
      RV log jitter H15 & & $\mathcal{U}(-5,5)$ & $-0$.03 $\pm$ 0.13\\
      RV log jitter HN & & $\mathcal{U}(-5,5)$ & 0.152 $\pm$ 0.88\\
      RV log jitter E19 & & $\mathcal{U}(-5,5)$ & $-0$.18 $\pm$ 0.13\\
      K1$\rm_{cycle}$ FWHM & m\,s$^{-1}$ & $\mathcal{U}(0,3\sigma FWHM)$ & 3.23 $\pm$ 0.88\\
      K2$\rm_{cycle}$ FWHM & m\,s$^{-1}$ & $\mathcal{U}(0,3\sigma FWHM)$ & 0.91 $_{-0.60}^{+0.79}$\\
      FWHM offset H03 & m\,s$^{-1}$ & $\mathcal{N}(0,3\sigma FWHM)$& $-1$.05$_{-0.72}^{+0.74}$ 2.4 \\
      FWHM offset H15 & m\,s$^{-1}$ & $\mathcal{N}(0,3\sigma FWHM)$ & 0.1 $\pm$ 1.0 \\
      FWHM offset HN 1& m\,s$^{-1}$ & $\mathcal{N}(0,3\sigma FWHM)$ & 1.83$_{-0.79}^{+0.83}$\\
      FWHM offset HN 2& \si{\meter\per\second} & $\mathcal{N}(0,3\sigma FWHM)$ & $-1$0.6 $\pm$ 2.0\\
      FWHM offset E19 & \,m\,s$^{-1}$ & $\mathcal{N}(0,3\sigma FWHM)$ & $-1$.6$_{-2.2}^{+2.1}$\\ 
      FWHM log jitter H03 & & $\mathcal{U}(-5,5)$ & 0.69 $_{-0.11}^{+0.12}$\\
      FWHM log jitter H15 & & $\mathcal{U}(-5,5)$ & 1.17 $\pm$ 0.12 \\
      FWHM log jitter HN & & $\mathcal{U}(-5,5)$ &  0.65 $\pm$ 0.11\\
      FWHM log jitter E19 & & $\mathcal{U}(-5,5)$ &  0.56 $_{-0.22}^{+0.18}$\\
      $K1_{\mathrm{cycle}}\, \rm S\,index\times 1000$ &  & $\mathcal{U}(0,3\sigma S\, index\times1000)$ & 5.9 $\pm$ 1.1\\
      $K2_{\mathrm{cycle}}\,\rm S\,index\times 1000$ &  & $\mathcal{U}(0,3\sigma S\,index\times1000)$ & 2.6 $_{-0.8}^{+1.0}$\\
      $\rm S\,index\times 1000$ offset H03 &  & $\mathcal{N}(0,3\sigma S\,index)$ & $-1$.74$_{-0.94}^{+0.95}$\\
      $\rm S\,index\times 1000$ offset H15 &  & $\mathcal{N}(0,3\sigma S\,index)$ & $-0$.8 $_{-1.3}^{+1.4}$\\
      $\rm S\,index\times 1000$ offset HN &  & $\mathcal{N}(0,3\sigma S\,index)$ & 2.2 $\pm$ 1.0\\
      $\rm S\,index\times 1000$ offset E19 &  & $\mathcal{N}(0,3\sigma S\,index)$ & $-2$.0$_{-2.9}^{+2.8}$\\
      $\rm S\,index\times 1000$ log jitter H03 & & $\mathcal{U}(-10,5)$ & $-0$.5$_{-2.7}^{+0.6}$ \\
      $\rm S\,index\times 1000$ log jitter H15 & & $\mathcal{U}(-10,5)$ & 1.15$_{-0.17}^{+0.15}$\\
      $\rm S\,index\times 1000$ log jitter HN & & $\mathcal{U}(-10,5)$ &
      0.15$_{-0.20}^{+0.18}$\\
      $\rm S\,index\times 1000$ log jitter E19 & & $\mathcal{U}(-10,5)$ & 0.89 $\pm$ 0.14\\
      Log A1 $S$ index $\times$ 1000 & & $\mathcal{U}(-5,5)$& 1.18$_{-0.11}^{+0.12}$\\
      $P\rm _{rot}$ & \si{\day} & $\mathcal{U}(15,45)$& 28.22$_{-0.30}^{+0.35}$\\
      log Timescale & & $\mathcal{U}(3,7)$& 4.68 $_{-0.23}^{+0.22}$\\
      log $\eta$ & & $\mathcal{U}(-5,2)$& $-$0.52 $\pm$ 0.14\\
      A1 RV & m\,s$^{-1}$ & $\mathcal{U}(-5,5)$& 1.02$_{-0.016}^{+0.018}$\\
      A1 FWHM & & $\mathcal{U}(-5,5)$ & 3.43$_{-0.37}^{+0.44}$\\
      A2 RV & \si{\meter} & $\mathcal{U}(-20,20)$& 2.64$_{-0.65}^{+0.74}$\\
      A2 FWHM & \si{\meter} & $\mathcal{U}(-20,20)$& 3.2 $\pm$ 1.4 \\
      \bottomrule
    \end{tabular}
    \medskip 
    
    \begin{minipage}{0.5\textwidth}
        \raggedright
    \end{minipage}
  \end{threeparttable}
\end{table*}

\begin{table}[]
  \centering
  \caption{RMS of the residuals. }
    \label{residuals_rms_table}
    \begin{tabular}{l r r r r r}
    \hline
    \hline
    \noalign{\smallskip}
    Model RMS &  Total & H03 & H15 & HN & E19 \\
    \noalign{\smallskip}
    \hline
    \noalign{\smallskip}
      RV - offset (\si{\meter\per\second})  & 1.77 & 2.17 & 1.41 & 1.51 & 1.28 \\
      RV - offset - stellar activity (\si{\meter\per\second}) & 1.42 & 1.73 & 1.17 & 1.34 & 0.96 \\
      RV - offset - stellar activity - HD 161098 b (\si{\meter\per\second}) & 1.32 & 1.63 & 1.06 & 1.31 & 0.83 \\
      RV - offset - stellar activity - HD 161098 b - HD 161098 c (\si{\meter\per\second}) & 1.22 & 1.49 & 1.01 & 1.21 & 0.77 \\
      \bottomrule
    \end{tabular}
\end{table}

\begin{figure*}[]
    \centering
    \includegraphics[width=1\linewidth]{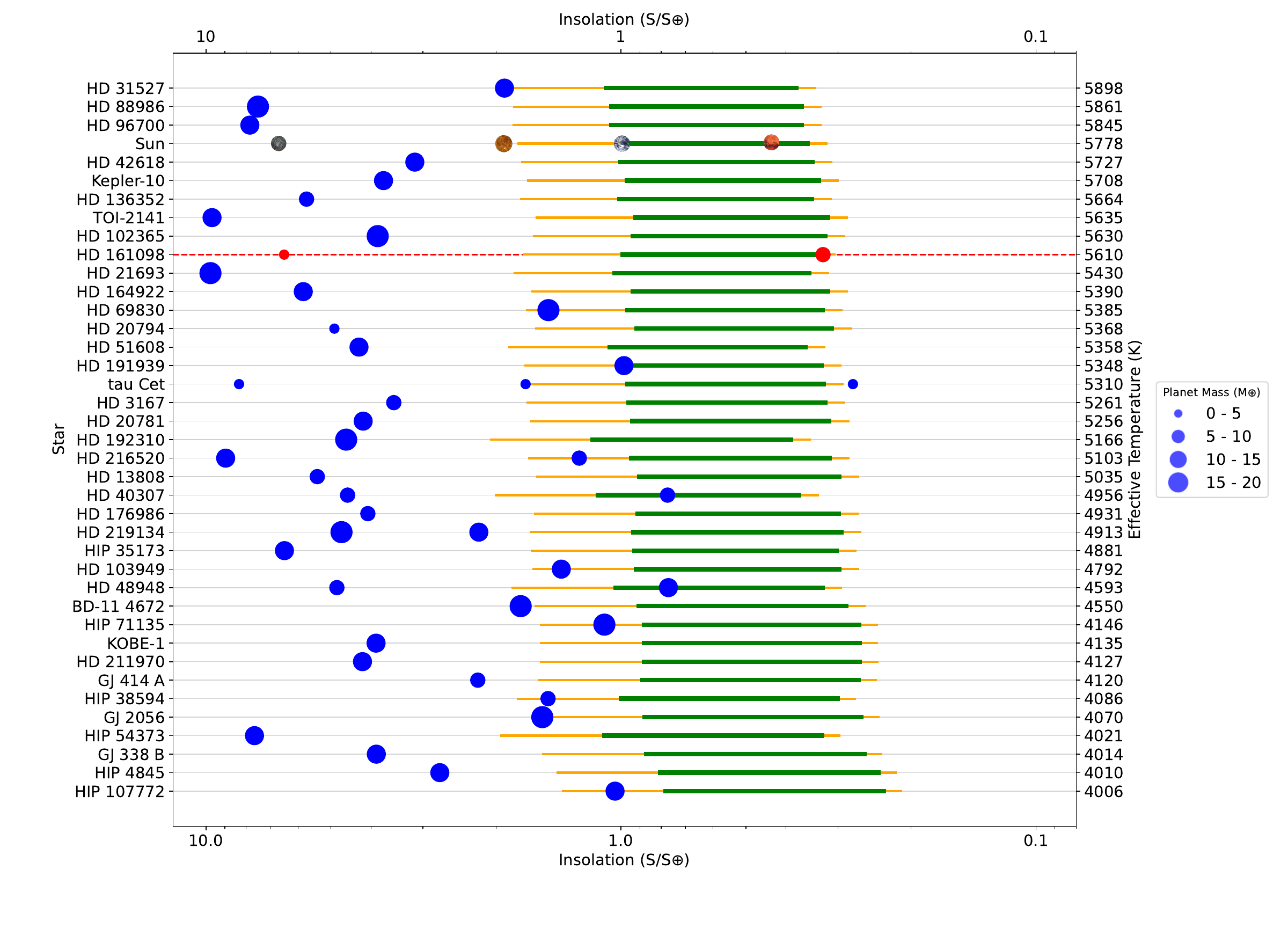}
    \caption{Comparison between the candidate system of HD 161098 and exoplanetary systems present in the literature for stars hotter than 4000K. We limit ourselves to the comparison with G and K-type stars because this is the scope of the RoPES survey. Only 3 exoplanets have been discovered orbiting the conservative HZ of their parent star (green region), and 13 planets have been discovered orbiting the optimistic HZ (yellow region). The parameters of the presented planets are from the NASA Exoplanet Archive \citep{2025_christiansen_archive}.}
    \label{fig_literature_hd161098}
\end{figure*}

\end{appendix}
\end{document}